%% file: 0_MAIN.tex
\documentclass[sn-nature,margin=1.5cm]{sn-jnl}

\usepackage{graphicx}%
\usepackage{multirow}%
\usepackage{amsmath,amssymb,amsfonts}%
\usepackage{amsthm}%
\usepackage{mathrsfs}%
\usepackage[title]{appendix}%
\usepackage{xcolor}%
\usepackage{textcomp}%
\usepackage{manyfoot}%
\usepackage{booktabs}%
\usepackage{algorithm}%
\usepackage{algorithmicx}%
\usepackage{algpseudocode}%
\usepackage{listings}%
\usepackage{caption}
\usepackage{comment}
\usepackage{array}
\usepackage[version=4]{mhchem} 
\usepackage{siunitx} 
\usepackage{amsmath,amssymb,gensymb,amsfonts,latexsym,mathtools} 
\usepackage{empheq,xfrac} 
\usepackage{nameref}
\newcommand*{\supsecref}[1]{\hyperref[{#1}]{Supp. Section: \nameref*{#1}}}
\theoremstyle{thmstyleone}%
\theoremstyle{thmstyletwo}%

\theoremstyle{thmstylethree}%

\begin{document}

\title[Article Title]{Unraveling energy flow mechanisms in semiconductors by ultrafast spectroscopy: Germanium as a case study}

\author[1]{\fnm{Grazia} \sur{Raciti}} 
\equalcont{These authors contributed equally to this work.}

\author*[1]{\fnm{Begoña} \sur{Abad}}
\email{b.abad@unibas.ch}
\equalcont{These authors contributed equally to this work.}

\author[2]{\fnm{Riccardo} \sur{Dettori}}

\author[3]{\fnm{Raja} \sur{Sen}}

\author[1]{\fnm{Aswathi} \sur{K. Sivan}} 

\author[1]{\fnm{Jose M.} \sur{Sojo-Gordillo}} 

\author[4]{\fnm{Nathalie} \sur{Vast}}

\author[5]{\fnm{Riccardo} \sur{Rurali}}

\author[2]{\fnm{Claudio} \sur{Melis}}

\author[4]{\fnm{Jelena} \sur{Sjakste}}

\author*[1]{\fnm{Ilaria} \sur{Zardo}} \email{ilaria.zardo@unibas.ch}

\affil[1]{\orgdiv{Department of Physics}, \orgname{University of Basel}, \orgaddress{\city{Basel}, \postcode{4056}, \country{Switzerland}}}

\affil[2]{\orgdiv{Department of Physics}, \orgname{University of Cagliari}, \orgaddress{\city{Monserrato}, \postcode{CA 09042}, \country{Italy}}}

\affil[3]{\orgdiv{SATIE, CNRS, ENS Paris-Saclay}, \orgname{Université Paris-Saclay}, \orgaddress{\city{ Gif-sur-Yvette}, \postcode{91190}, \country{France}}}

\affil[4]{\orgdiv{Laboratoire des Solides Irradi\'es}, \orgname{CEA/DRF/IRAMIS, Ecole Polytechnique, CNRS, Institut Polytechnique de Paris}, \orgaddress{\city{Palaiseau}, \postcode{91128}, \country{France}}}

\affil[5]{\orgname{Institut de Ci\`encia de Materials de Barcelona, ICMAB--CSIC}, \orgaddress{\street{Campus UAB}, \city{Bellaterra}, \postcode{08193}, \country{Spain}}}


\abstract{
Semiconductor materials are the foundation of modern electronics, and their functionality is dictated by the interactions between fundamental excitations occurring on (sub-)picosecond timescales. Using time-resolved Raman spectroscopy and transient reflectivity measurements, we shed light on the ultrafast dynamics in germanium. We observe an increase in the optical phonon temperature in the first few picoseconds, driven by the energy transfer from photoexcited holes, and the subsequent decay into acoustic phonons through anharmonic coupling. Moreover, the temperature, Raman frequency, and linewidth of this phonon mode show strikingly different decay dynamics. This difference was ascribed to the local thermal strain generated by the ultrafast excitation. We also observe Brillouin oscillations, given by a strain pulse traveling through germanium, whose damping is correlated to the optical phonon mode. These findings, supported by density functional theory and molecular dynamics simulations, provide a better understanding of the energy dissipation mechanisms in semiconductors.
}




\maketitle



\section{Introduction}\label{Intro}
Semiconductor devices are the building blocks of most modern technologies. Their speed, efficiency, and performance are directly related to the energy transfer processes during external excitations. More specifically, the charge and heat carrier dynamics and their mutual interaction are the fundamental mechanisms that govern their macroscopic physical properties and, consequently, set limits to their operation conditions. Therefore, a fundamental understanding of the energy decay channels after excitation enables improved device performance \cite{Hathwar_2019}, advancing heat management of next-generation devices such as solar cells and optical detectors. In particular, germanium (Ge) is a key material for the semiconductor industry, with applications spanning a wide range, from electronics \cite{goley2014, Arianna2024} to photonics \cite{Soref2010} to quantum technologies \cite{Scappucci2020}. Ge is a group IV semiconductor with an indirect bandgap, which attracts much attention due to its high carrier mobility \cite{Rossner2004} and compatibility with silicon-based systems. Although this material has been extensively researched and many of its properties have been well characterized for decades, the fundamental mechanisms governing energy dissipation in Ge are still poorly understood. Indeed, to date, only a handful of studies exploring the carrier and phonon dynamics after ultrafast excitation in Ge are available\cite{Young1989, othonos_probing_1998, Zurch2017, Sjakste:2025} and cannot fully describe the mechanisms behind it. Technological and computational progresses enable now to probe dynamics with an unprecedented time and energy resolution. 

In particular, carrier and phonon dynamics in Ge have been studied by time-resolved Raman spectroscopy (TRRS), a pump-probe spectroscopy technique that probes incoherent phonon lifetimes \cite{prasankumar_optical_2012}. These studies are mainly divided into low and high photoexcited carrier densities \text{n}, namely \text{n} $ <10^{20} \, \text{cm}^{-3}$ and \text{n} $ >10^{20} \, \text{cm}^{-3}$, respectively. In general, after ultrafast pump excitation, electrons (holes) will initially be excited to high-energy states in a timescale of femtoseconds. These carriers redistribute energy among themselves via carrier-carrier scattering in tens of femtoseconds, forming a hot Fermi-Dirac distribution. These hot electrons (holes), which have a mean kinetic energy considerably higher than the thermal energy of the lattice, will relax to the bottom (top) of the conduction (valence) band by emitting optical and acoustic phonons in a few picoseconds. Subsequently, they reach thermal equilibrium with the lattice in about \qtyrange{2}{100}{ps}, after which carrier recombination occurs \cite{shah_ultrafast_1999, Othonons_semiconductors}. 

At moderate photoexcited hole densities, Young \textit{et al.} studied the lifetime of the non-equilibrium longitudinal (LO) and transversal (TO) optical phonons in Ge. Both LO and TO phonons were shown to have a lifetime of \qty{4}{ps} at room temperature. The fact that both phonon populations were similar was ascribed to the generation of phonons by heavy holes in the vicinity of the zone center \cite{young_picosecond_nodate}. A year later, Othonos \textit{et al.} used a combination of TRRS and transient reflectivity (TR) to study the hot-carrier dynamics in Ge \cite{othonos_hot-carrier_1989}. Both works found that non-equilibrium optical phonon populations in Ge are effectively generated and probed by TRRS. At high photoexcited hole densities, Ledgerwood \textit{et al.} found that phonon reabsorption by holes is responsible for faster optical phonon decay times ($<$\qty{3}{ps}) than previously demonstrated for lower photoexcited densities \cite{ledgerwood_enhanced_1994, ledgerwood_picosecond_1996}. 

However, the time resolution of these studies was limited by the pulse duration of \qty{4}{ps}, so only phonon lifetimes longer than this pulse duration could be probed. Shorter lifetimes at higher photoexcited carrier densities were extracted from the linewidth broadening observed during time-averaged anti-Stokes measurements. In this technique, Raman spectra are probed with a picosecond probe pulse without pump excitation \cite{ledgerwood_picosecond_1996}. The lifetime extracted from this broadening may not be strictly comparable to the lifetime measured by time-resolved measurements, as time-averaging does not follow the phonon population after a single excitation. Therefore, more precise time-resolved Raman spectroscopy measurements with pump and probe pulse durations shorter than the expected phonon lifetime are essential for understanding phonon dynamics.  \\

Raman spectroscopy also enables the study of the phonon self-energy, i.e., phonon frequency and lineshape broadening. In group IV semiconductors, such as Ge, the interaction between carriers and phonons takes place through the deformation potential \cite{Klenner_1992}. This mechanism may lead to a contribution of the photoexcited carriers to the phonon self-energy \cite{ledgerwood_picosecond_1996}. In particular, at high photoexcited hole densities, single-particle excitations may play a role in the renormalization of the phonon energy, similarly to the line broadening and frequency shift in highly p-doped Ge as compared to intrinsic Ge \cite{olego_self-energy_1981}. On the other hand, pump-induced strain may also affect the phonon self-energy \cite{Cerdeira_1971}. However, extraordinary advances in ultrafast laser and detection technology as well as progress in computational techniques allow now to experimentally address the dynamics of the phonon self-energy and phonon temperature in Ge with improved accuracy and temporal resolution. Therefore, a systematic study of the ultrafast dynamics of energy dissipation channels in Ge is needed to shed light on the fundamental mechanisms that will enable improved devices. Understanding how these processes evolve is critical for the optimal design of Ge-based semiconductor nanocomponents.   \\

In this work, we employ TRRS and TR to study the energy dissipation dynamics following ultrafast laser excitation on a $(111)$-oriented bulk Ge sample. In particular, we study the optical phonon population and its temperature dynamics within the first few picoseconds after ultrafast excitation, observing both the rise and its subsequent decay. We extract the relaxation decay time for all Raman spectral features -- i.e., intensity, Raman frequency, and linewidth -- and compare our experimental results with molecular dynamics (MD) and density functional theory (DFT), finding excellent agreement. Finally, we probe the correlation between optical phonon modes and coherent acoustic waves excited in the system.


\section{Results}
\subsection{Phonon dynamics in germanium}\label{sec2.1}
\begin{figure}
\begin{center}
\includegraphics[width=\textwidth]{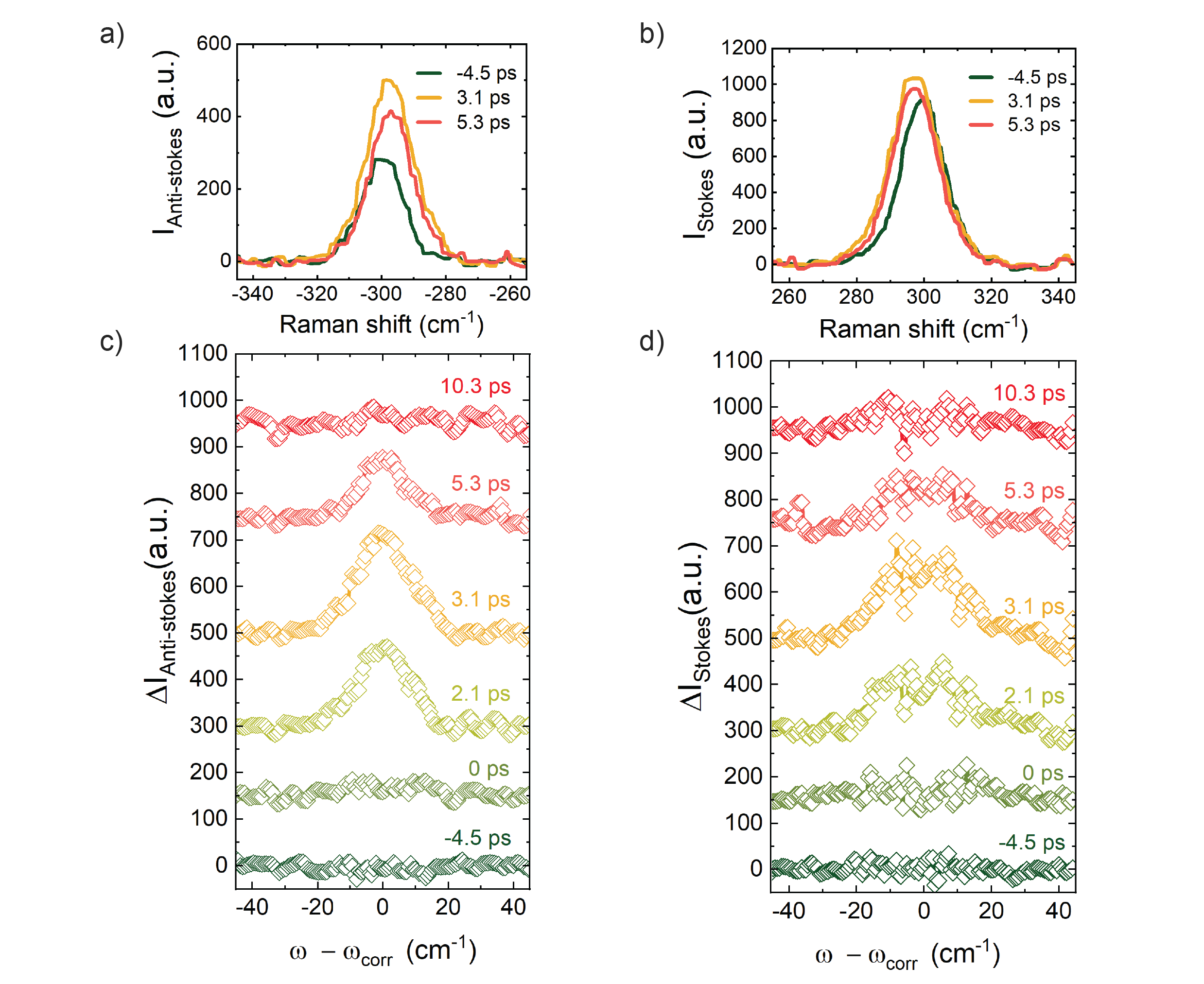}
\caption{\textbf{Time-resolved spontaneous Raman spectroscopy}. Raman signal of the germanium TO/LO phonon mode detected at selected delay times before and after pump excitation at a carrier density of \qty{1.5e20}{cm^{-3}}. Upper panels: Raman spectra for three representative times showing both the change in intensity and phonon frequency for the anti-Stokes (a) and Stokes (b) bands. Negative (positive) times correspond to spectra measurements before (after) the arrival of the pump pulse. Lower panels: difference Raman spectra for anti-Stokes (c) and Stokes (d) scattering, extracted by subtracting the average spectra taken at negative delay times. To account for spectral drift as a function of time, prior to the subtraction, each spectrum is offset by $\Delta \omega_{corr}$, whih is the difference between its peak position and that of the average Raman shift of all spectra before excitation.}
\label{fig:Intensity}
\end{center}
\end{figure}

We perform TRRS (described in \nameref{sec4}) on an intrinsic crystalline Ge sample oriented along the $\langle 111 \rangle$ direction. In brief, an ultrafast pump laser pulse brings the system out of equilibrium, and a subsequent probe pulse monitors the evolution of the Raman spectrum as a function of delay time. Therefore, we track the dynamics of the TO/LO phonon mode of Ge (see \nameref{sec4} and Supplementary \ref{Exp_details} for further details of the experimental setup, and experimental conditions). As Raman spectroscopy probes phonon modes near the $\Gamma$-point, TRRS is particularly well suited to explore phonon dynamics in Ge, since the energy deposited in carriers after ultrafast excitation flows preferentially to low wavevectors due to the conservation of energy and momentum \cite{Young1989}. \autoref{fig:Intensity}a~and~b show the TO/LO mode intensity, located at $\pm$~ \qty{300.7}{cm^{-1}} \cite{Parker_1967}, of the anti-Stokes (AS), and Stokes (S) Raman scattering, respectively, at various representative times for an excited carrier density of \qty{1.5e20}{cm^{-3}}. Although both bands undergo an increase in intensity with time, the AS scattering shows a larger change in relative intensity ($\Delta I/I$) than that of the S scattering, gaining up to \qty{45}{\%} of intensity at \qty{3.5}{ps}. Instead, the maximum change in intensity for the S band is only \qty{13}{\%} at the same delay time. This is expected from the dependence of the S and AS scattering on the phonon population, which is $n+1$ and $n$, respectively \cite{prasankumar_optical_2012}. Although the AS scattering is less likely to occur at equilibrium, i.e., near room temperature, because it requires pre-existing excited phonons, the ultrafast excitation increases the population of high-energy phonons, causing a larger relative increase in the AS intensity as compared to the S intensity. Moreover, \autoref{fig:Intensity}a~and~b also show a clear red shift for both bands as a function of time. \\
To better visualize the change in intensity as a function of time, \autoref{fig:Intensity}c~and~d show the difference Raman spectra at various delay times before and after pump excitation for both AS and S, respectively (see Supplementary \ref{TRRS_processing} for further details on this subtraction). Despite $\Delta I/I$ being very different for both bands, their dynamics are very similar: there is an initial increase in intensity in the vicinity of the optical phonon response located at $\pm$ \qty{301}{cm^{-1}}, which reaches a maximum at \qty{3.5}{ps} and eventually decreases until it fully vanishes for $t>$\qty{10}{ps}. 

\begin{figure}[H]
\begin{center}
\includegraphics[width=\textwidth]{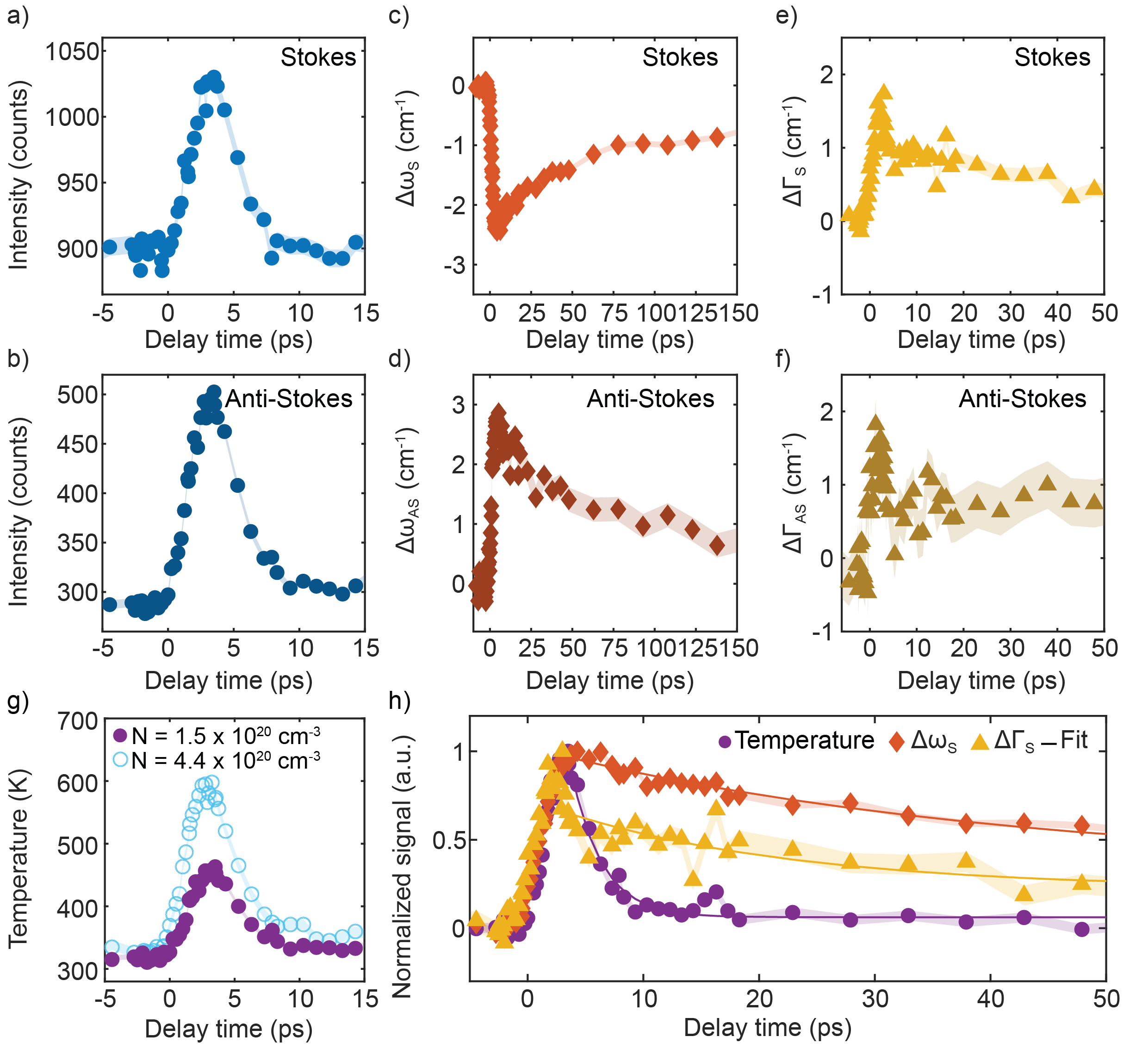}
\caption{\textbf{Dynamics of the Ge TO/LO phonon mode}. Spectral features are extracted by fitting the Raman peak to a Gaussian fit at the acquired delay times. a-b)~Intensity extracted from the Gaussian fit of Stokes (a) and Anti-Stokes (b). c-d)~ Change in Raman frequency shift ($\Delta \omega$) of Stokes (c) and Anti-Stokes (d). e-f)~Change in linewidth ($\Delta \Gamma$) extracted for Stokes (e) and Anti-Stokes (f) scattering, respectively. g) Temperature of the phonon mode calculated from the anti-Stokes/Stokes intensity ratio for low and high excitation carrier densities: \qty{1.5e20}{cm^{-3}} and \qty{4.4e20}{cm^{-3}}, respectively. There is a rise corresponding with the building up of the TO/LO mode, followed by a decay with a time constant of \qty{2.7(0.2)}{ps} and \qty{2.3(0.1)}{ps} for the low and high fluences, respectively. h)~Comparison of the normalized dynamics of the temperature, Raman frequency $\Delta \omega_S$, and linewidth $\Delta\Gamma_S$ of the Stokes scattering for the low fluence. The three signals have a similar rise while their decay times are strikingly different, indicating the contribution of nonthermal mechanisms to the frequency and linewidth decays of the TO/LO phonon mode. In all plots, the estimated error is represented by the shading.}
\label{fig:PhonProp}
\end{center}
\end{figure}

To perform a more quantitative analysis of the dynamics of the spectral features, the S and AS Raman spectra are fitted with a Gaussian profile in a spectral region of $\pm$ \qtyrange{260}{340}{cm^{-1}}, from which the peak intensity, Raman frequency, and linewidth, i.e. full width at half maximum (FWHM), are extracted. The fitted peak intensity, change in phonon frequency ($\Delta \omega$), and change in linewidth ($\Delta \Gamma$) of both S and AS Raman scattering are shown in \autoref{fig:PhonProp}, where the shading corresponds to the estimated experimental error based on the standard deviation of all spectra acquired prior to the pump excitation. 

The temporal evolution of the intensities of the S and AS peaks is shown in \autoref{fig:PhonProp}a~and~b, respectively. An intensity rise reaches its maximum at $\sim$\qty{4}{ps}, after which the intensity decays and fully returns to equilibrium values after $\sim$\qty{10}{ps}. 
In contrast, $\Delta \omega(t)$ of both S and AS, shows completely different dynamics than that of the intensity (see \autoref{fig:PhonProp}c~and~d, respectively). The rise in its absolute value, $\lvert\Delta \omega \rvert$, still follows dynamics similar to that of the intensity, which also takes $\sim$\qty{4}{ps} to reach its maximum value for both bands. Nevertheless, the most notable difference is its slower decay, where $\Delta \omega(t)$ does not fully decay back to its equilibrium value within the experimental temporal window (\qty{200}{ps}). \autoref{fig:PhonProp}e~and~f show the dynamics of $\Delta \Gamma$, for S and AS Raman scattering, respectively. Similarly to the intensity and Raman shift, $\Delta \Gamma$ also reaches its maximum value of approximately $\sim$\qty{1.5}{cm^{-1}} in $\sim$\qty{4}{ps} for both bands. However, its decay is strikingly different from that of both the intensity and the $\Delta \omega(t)$.

The S and AS peak intensities provide valuable information, as one can calculate the TO/LO phonon temperature from their ratio (see Supplementary \ref{Ph_temp} for detailed explanation). \autoref{fig:PhonProp}g shows the temperature evolution of the TO/LO Ge phonon mode, before and after excitation, for low (\qty{1.5e20}{cm^{-3}}) and high pump fluences (\qty{4.4e20}{cm^{-3}}). For both cases, the temperature at negative times, i.e., before pump excitation, is slightly above room temperature, indicating a \qty{20}{K} heating caused by the probe pulse during our measurements. The fluence of the probe is kept as low as possible to minimize this effect, while still providing a sufficiently large signal-to-noise ratio. After pump excitation, there is an increase in temperature following the dynamics of the AS intensity, reaching its maximum at $\sim$\qty{4}{ps}, and subsequently decaying back to equilibrium with a decay time of \qty{2.7(0.2)}{ps} and \qty{2.3(0.1)}{ps} for low and high fluences, respectively, extracted by fitting the decay using a single exponential decay function for both curves. Interestingly, we observe a slightly faster decay as we increase fluence, possibly indicating the presence of an additional energy dissipation channel for optical phonons at larger fluences, i.e., phonon reabsorption by photoexcited carriers \cite{ledgerwood_enhanced_1994} \cite{ledgerwood_picosecond_1996}. \autoref{fig:PhonProp}h shows the comparison between the normalized temperature, phonon frequency and linewidth, showing noticeable different decay dynamics.

\begin{figure}
\begin{center}
\includegraphics[width=\linewidth]{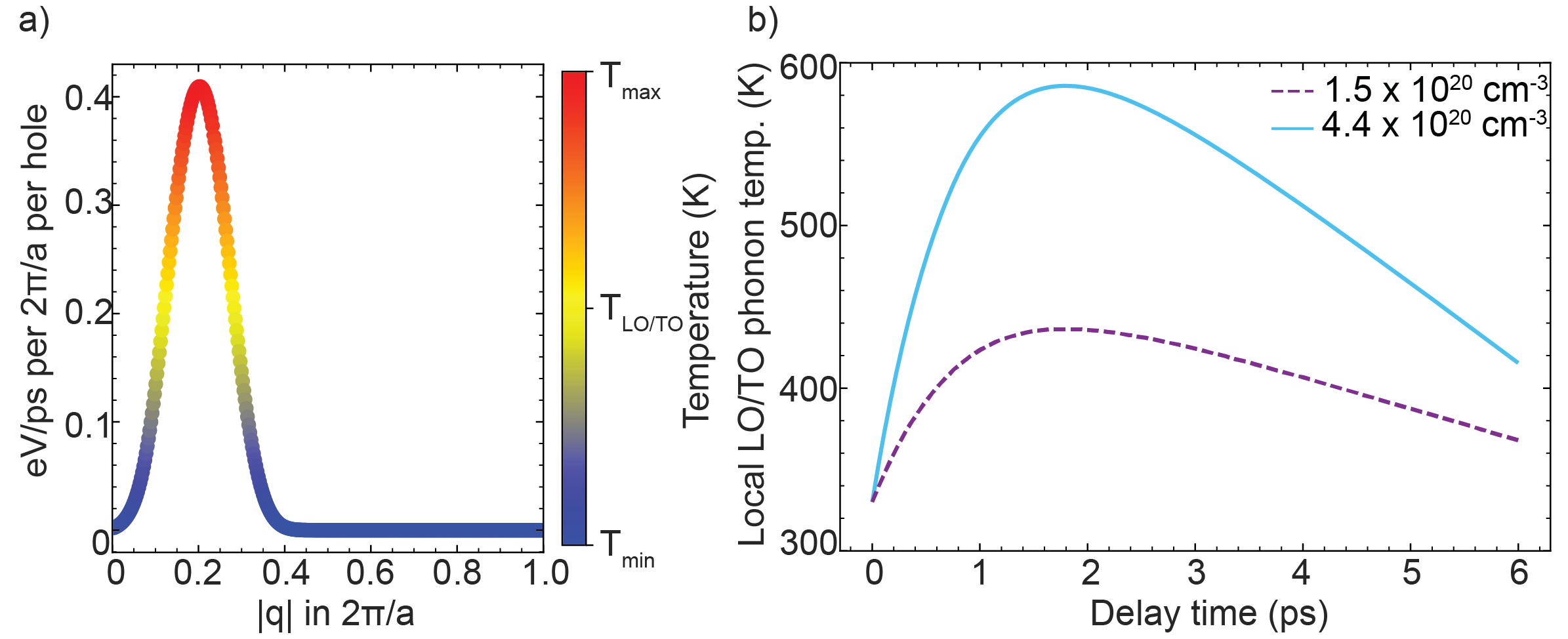}
\caption{Dynamics of temperature rise of TO/LO modes calculated with time-dependent two-temperature model for carriers and phonons, coupled with DFT-based description of the electron-phonon and phonon-phonon coupling in Ge (See Supplementary \ref{DFT}). a)~Hole-phonon energy transfer rate calculated with DFT as a function of phonon wavevector ($|\mathbf{q}|$-dependent spectral function, see Supplementary \ref{sec_DFT_energy_transfer}), which illustrates the localization in the Brillouin zone of the overheated optical modes. b)~Calculated local temperature rise of Raman optical modes due to hole-phonon interaction, for hole densities of \qty{1.5e20}{cm^{-3}} (dashed purple line) and \qty{4.4e20}{cm^{-3}} (light blue solid line). Subsequent decay into acoustical phonons is modeled with a calculated phonon-phonon decay rate of 2.5~ps (see text and Supplementary \ref{DFT}).}
\label{fig:DFT}
\end{center}
\end{figure} 

To understand the mechanisms behind the rise and subsequent decay of the TO/LO Raman-active phonon population and the temperature, we perform DFT calculations of the carrier-phonon and phonon-phonon interactions. The timescale of the temperature is determined by the relaxation dynamics of photoexcited carriers. According to our calculations, this rise is attributed to the photoexcited hole dynamics: the dominant scattering process for photoexcited holes is intravalley scattering with optical phonons, that leads to the heating of Raman-active optical modes (see Supplementary \ref{DFT} for full details of the calculations). In contrast, the dynamics of photoexcited electrons are mostly determined by the intervalley scattering \cite{Sjakste:2025} (see Supplementary \ref{sec:hotholes}), which explains why photoexcited electrons do not influence the dynamics of Raman-active optical modes. As seen in \autoref{fig:DFT}a, photoexcited holes transfer their energy only to phonons localized close to the Brillouin zone center, producing a strong rise in the local phonon temperature.
On the other hand, \autoref{fig:DFT}b shows our calculated phonon temperature dynamics. According to our DFT calculations, the observed decay times of \qty{2.7(0.2)}{ps} and \qty{2.3(0.1)}{ps} for high and low fluence, respectively, correspond to the decay of optical modes into acoustic modes via anharmonic coupling. Indeed, the lifetime of $\Gamma$-point optical phonons, calculated within the Relaxation Time Approximation and considering third-order anharmonic processes, is found to be between 2 and \qty{2.5}{ps} at room temperature (depending on what DFT and Boltzmann transport equation (BTE) solvers were used; see Supplementary \ref{DFT}), which is in very good agreement with the measured values.
Interestingly, the decay of optical modes into acoustic modes determined both experimentally and theoretically in our work is found to be faster than the one previously reported by Young \textit{et al.}  ($\sim$\qty{4}{ps} at room temperature \cite{young_picosecond_nodate}). However, it is worth noting that the temporal resolution of Young et al. \cite{young_picosecond_nodate} was limited by their pulse duration, \qty{4}{ps}, which makes it challenging to observe any faster phenomena. Moreover, our measured and calculated lifetime is in agreement with Mendez \textit{et al.}\cite{Menendez:1984} who reported a broadening of \qty{2}{cm^{-1}} for pristine Ge at \qty{300}{K}, which corresponds to the lifetime of \qty{2.5}{ps}.

Besides the S and AS intensity ratio, one can also extract temperature from the Raman frequency shift or the linewidth \cite{sandell_thermoreflectance_2020}. Interestingly, \autoref{fig:PhonProp}h shows that the temperature rise calculated from the intensity ratio, $\Delta\omega$ and $\Delta\Gamma$, are in very good agreement, which indicates that the contribution to the rise of the Raman frequency and linewidth is given exclusively by the increase of the phonon temperature. Indeed, the increase in temperature $\Delta T$ is about \qty{150}{K} at $\sim$\qty{4}{ps}, which corresponds to a $\Delta\omega$ and $\Delta\Gamma$ of \qty{2.4}{cm^{-1}} and \qty{1.5}{cm^{-1}}, respectively, as shown in \autoref{fig:PhonProp}c-e~and~d-f. These values are in excellent agreement with measurements from conventional Raman thermometry reported in previous works \cite{Menendez_1984, burke_temperature_1993}.  On the other hand, \autoref{fig:PhonProp}h shows that the normalized temperature, Raman shift, and linewidth have remarkably different decays, indicating that alongside with a decay in temperature, there are additional contributions to the decay dynamics of the Raman frequency and the linewidth. Particularly, we extract a temperature decay time (\qty{2.7 \pm 0.2}{ps}), that is drastically different from that of $\Delta\omega$ (\qty{44.5 \pm 3.7}{ps}) and $\Delta\Gamma$ (\qty{19.4 \pm 5.2}{ps}), shown in \autoref{tab:decays}. This indicates that there are more contributions to the decay of these magnitudes in addition to the change caused by the decrease in temperature. In general, the thermal contribution to the Raman shift is influenced by phonon-phonon scattering, i.e., anharmonicity, and the thermal expansion of the lattice caused by the change in temperature. However, the linewidth is usually only influenced by anharmonicity \cite{sandell_thermoreflectance_2020}. To explore what other phenomena can influence both $\Delta\omega$ and $\Delta\Gamma$, we performed TR experiments under the same conditions, as well as MD simulations.

\begin{table}[!h]
\begin{tabular}{ccccccc} 
 \hline
 $\tau_{T} (ps)$ & & & $\tau_{\Delta \omega(Stokes)} (ps)$ & & & $\tau_{\Delta \Gamma(Stokes)} (ps)$ \\  
 \hline
 $2.7\pm 0.2$ & & & $44.5\pm 3.7$ & & & $19.4\pm 5.2$  \\ 
 \hline 
\end{tabular}
\caption{Decay constants, $\tau$, extracted from a single exponential decay fit of the different spectral features, namely: the temperature extracted from the ratio between the anti-Stokes and Stokes intensities, $T$; the change in Raman shift, $\Delta \omega$, and the change in linewidth, $\Delta \Gamma$, of the Stokes band.}
\label{tab:decays}
\end{table}

\begin{figure}
\begin{center}
\includegraphics[width=\textwidth]{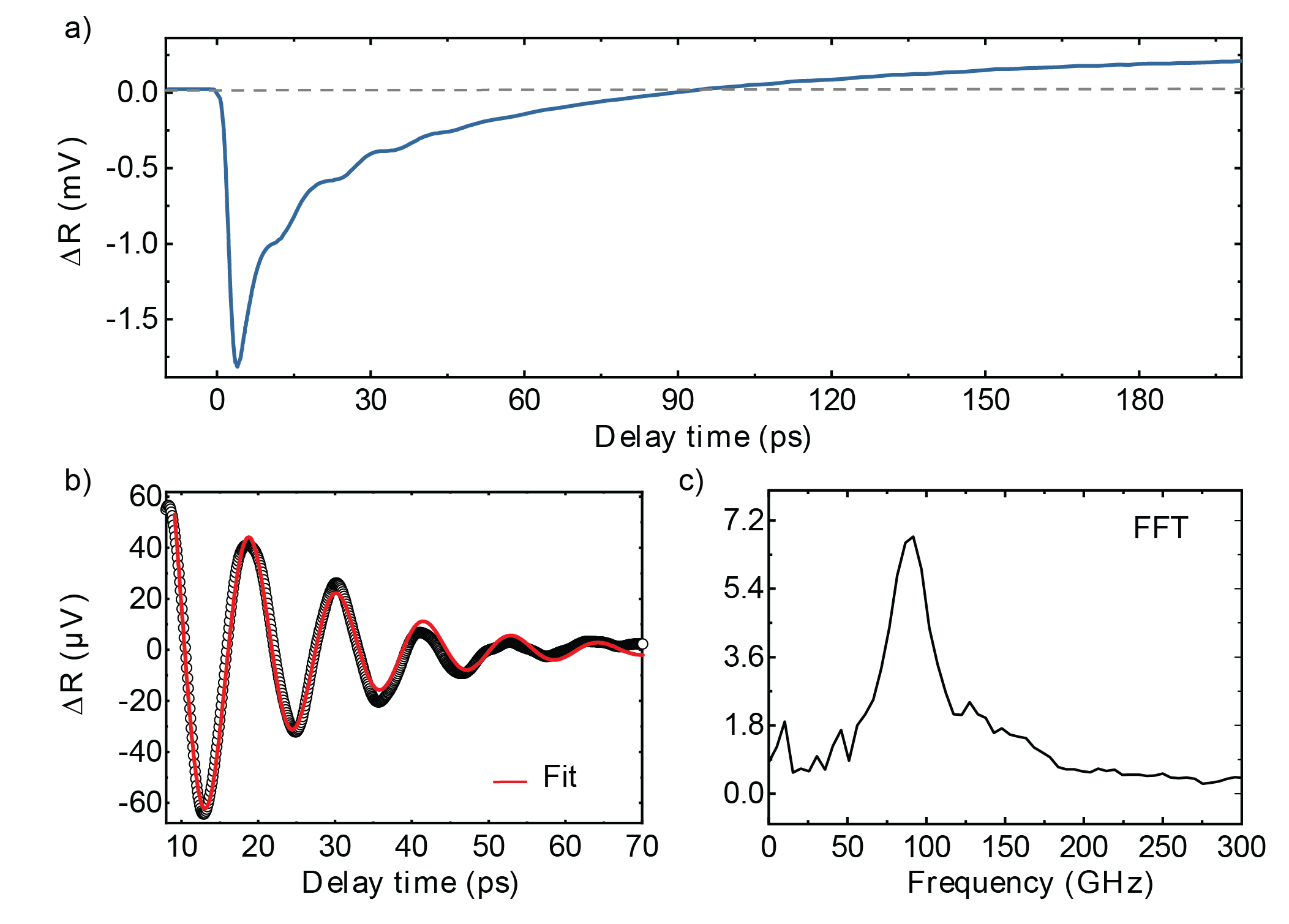}
\caption{\textbf{Transient reflectivity on Ge}. a)~Change in the reflectivity signal of bulk Ge as a function of time for an excitation of $1.4 \cdot 10^{20} \, \text{cm}^{-3}$, i.e., in the same conditions as the measurements performed by TRRS. The signal shows a sharp decrease in reflectivity after excitation, followed by a recovery eventually increasing above positive values. b)~Oscillating component of the signal after the subtraction of a tri-exponential decay following the initial decrease in reflectivity (empty circles) and its corresponding fit to a damped oscillating function (solid red line). c)~Frequency spectrum calculated by taking the Fourier transform of the oscillating signal.}
\label{fig:Reflectivity}
\end{center}
\end{figure}

The pump-probe setup used to perform TRRS can be adapted to measure TR under the same exact low fluence conditions (see \nameref{sec4} and Supplementary \ref{Exp_details} for further details of the setup). \autoref{fig:Reflectivity}a shows the change in reflectivity as a function of time after pump excitation of bulk germanium oriented along the $\langle 111 \rangle$ axis. This signal is related to the change in the material's dielectric constant, $\varepsilon$, caused by the carrier dynamics triggered after excitation. Among the factors that modify $\varepsilon$, there are two main contributors that can affect $\varepsilon$ in opposite ways, namely the charge density and the lattice expansion \cite{Wagner_1996}. 

The rise of the TR signal occurs approximately up to \qty{4}{ps}, as in the case of the Raman spectral properties shown in \autoref{fig:PhonProp}. After this rise, there is a subsequent decay, which we can fit to a tri-exponential decay with time constants of \qty{2.4(0.6)}{ps}, \qty{11.2(3.6)}{ps}, and \qty{68(14)}{ps}. The contributions involved to the dynamics cancel each other out and eventually make the TR signal rise from negative towards positive values, as shown by the horizontal dashed line in  \autoref{fig:Reflectivity}a. This is likely the result of two competing contributions. Initially, as the charge carriers population grows and then decays, they hinder the reflectivity signal, while in a second phase, the lattice contribution dominates with a positive contribution to the TR signal, flipping the signal. 

Noteworthy, the subtraction of the decay dynamics reveals a strong oscillating signal (see \autoref{fig:Reflectivity}b). We fit this signal with a damped oscillating function, from which we calculate a damping time constant of \qty{18.1(3.0)}{ps} and an oscillation period of \qty{11(1)}{ps}. This period corresponds to an oscillation frequency of \qty{91.4(1)}{GHz} as confirmed by the Fourier transform of this oscillating signal, whose frequency spectrum is shown in \autoref{fig:Reflectivity}c.

The origin of these oscillations lies in the generation of a coherent acoustic phonon wave packet (acoustic strain field) by the pump pulse, which then propagates through the sample \cite{rathore_long-lasting_2023,aagaard_measurement_2024}. We identify this signal modulation as Brillouin oscillations, which arise from the interaction between the light and the generated strain pulse traveling through the sample (see Supplementary \ref{sec:Brillouin_osc} for more details) \cite{ezzahri_coherent_2007}. Interestingly, the damping time constant of the oscillation is in agreement with the decay time of the linewidth measured by TRRS. This indicates that the strain pulse triggered by ultrafast excitation also influences the dynamics of the phonon linewidth. These coherent acoustic phonons are usually damped over time as phonons lose coherence, due to coupling with other particles, anharmonicity, or defects. Indeed, the Raman linewidth is a measurement of these phenomena, as the peak broadening is related to the phonon lifetime and damping mechanisms. 
\\

\subsection{Influence of strain on phonon dynamics}\label{sec2.2}

To understand more about the effect of strain on the phonon dynamics, we performed MD computational experiments of bulk Ge in a computational cell ($L=85.1$~nm) that we partitioned into three regions (Supporting Information for full details of the calculations). After equilibrating the system at \qty{300}{K}, we used a Generalized Langevin Thermostat (GLE)~\cite{CeriottiPRL09, CeriottiPCS10, DettoriJCTC17} that selectively pumps energy in the central region ($2.8$~nm-long) only on vibrational modes within a frequency window $\Delta \omega = 0.5$~cm$^{-1}$ centered around $\omega_0 = 310$~cm$^{-1}$. This selective thermalization of atomic degrees of freedom mimics the experimental conditions, where the laser injects energy into specific modes ({\it pump}). We maintain this nonequilibrium condition for \qty{500}{fs}, realizing two different excitation conditions at $T_{\rm exc}=450$~K and \qty{600}{K} --corresponding to the experimental maximum temperatures measured for the two explored fluences-- and then remove any temperature constraint, letting the system free to reach again the thermodynamic equilibrium at \qty{300}{K}. We monitor the decay of the phonon excited state by computing the time-dependent vibration density of states (vDOS) (see Supplementary \ref{sec:MD} for more details). In \autoref{fig:MD}a we show how the excess energy results in an overpopulated optical peak of the vibrational spectrum, located at around $\omega_0 = 310$~cm$^{-1}$, with decreasing intensity as a function of time. In fact, the energy, initially placed in the optical phonon modes, quickly decays following a two-times exponential decay with decay constants $\tau_{I1} = 2.93\pm0.07$~ps  and $\tau_{I2} = 11.4\pm1.1$~ps  when exciting the system at \qty{450}{K} ($\tau_{I1} = 2.82\pm0.04$~ps and $\tau_{I2} =11.7\pm1.1$~ps for \qty{600}{K}, see Supplementary \autoref{fig:kinetic_temp}). It is worth highlighting that instead the experimental data could be fitted with a single exponential decay. We understand this difference considering that TRRS probes the dynamics of specific phonon modes, while the MD computational experiments can also capture further changes in the local temperature and strain affecting the VDOS.\\
Our simulation suggests that this decay is associated with the excess of energy being transferred to the lower frequency acoustic modes through phonon-phonon scattering, in agreement with our DFT calculations. As shown in \autoref{fig:MD}b, while the optical peak decays (orange circles, left y-axis), the peak at $\omega\sim$~\qty{220}{cm^{-1}} (blue triangles, right y-axis) is characterized by a fast rise and a slower decay. A similar qualitative behavior is observed for the acoustic peak centered around \qty{150}{cm^{-1}}, see Supplementary \autoref{fig:220_area} and \autoref{fig:150_area}. The low frequency phonons rise time (of about 3 ps) matches the decay time of the optical one, clearly indicating anharmonicity as the main energy dissipation mechanism for the cooling of the TO/LO phonon mode. 

\begin{figure}
\begin{center}
\includegraphics[width=\linewidth]{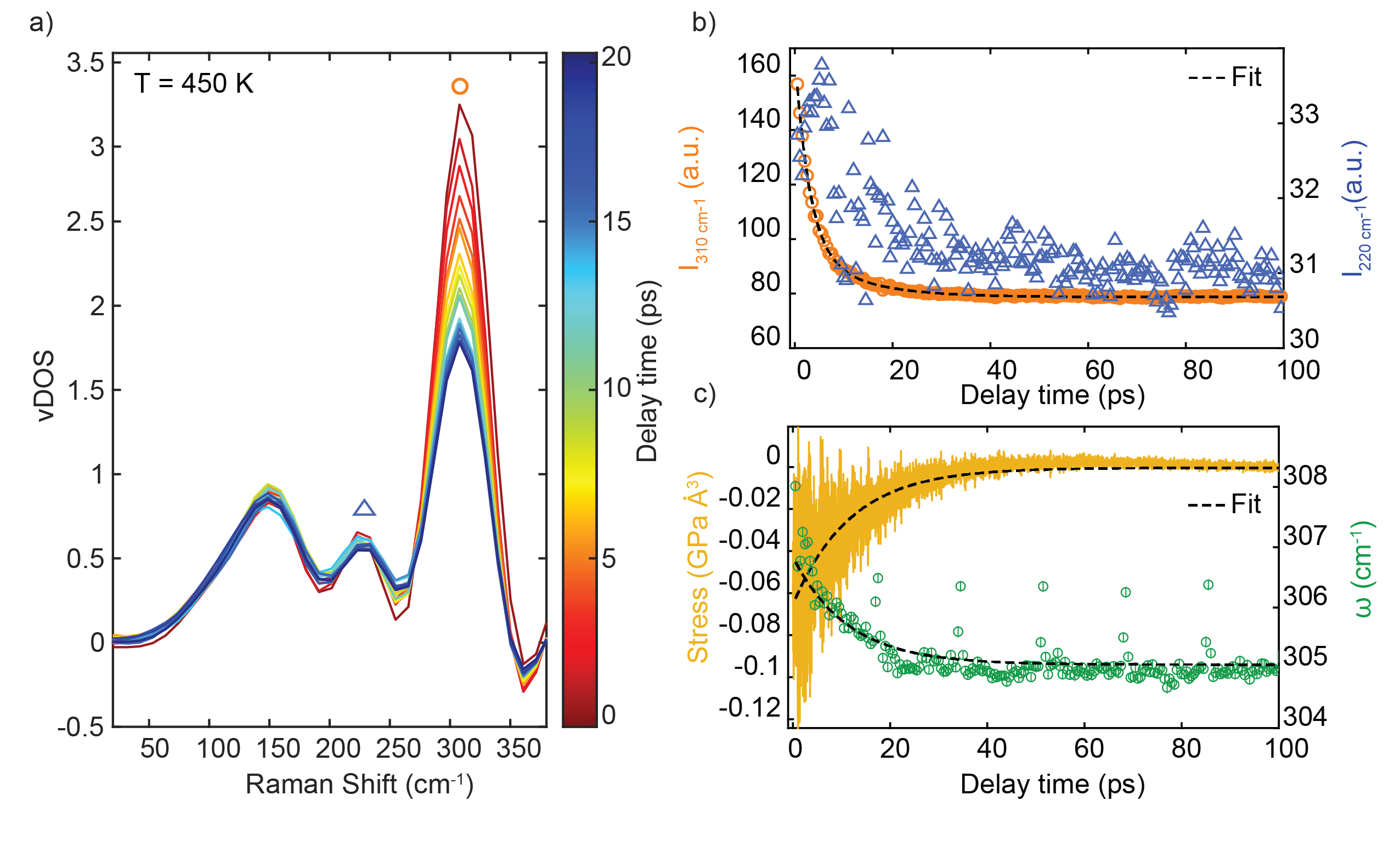}
\caption{{\textbf{Molecular dynamics results}. a)~VDOS at different delay times of the central region of the computational cell. The excess energy initially injected into modes around \qty{310}{cm^{-1}} (red curve) progressively decays and at sufficiently long times, the equilibrium VDOS is recovered (blue curve). b)~Computed intensity of the \qty{310}{cm^{-1}} peak marked with an orange dot in panel a (orange symbols, left y-axis; bi-exponential decay fit to the data as black dashed line) and of the \qty{220}{cm^{-1}} peak  marked with a blue triangle in panel a (blue triangles, right y-axis) of the VDOS as a function of time. (c)~Frequency of the optical mode (green circles, right y-axis) and stress (yellow line, left y-axis) as a function of time, together with the corresponding fitted exponential trend (single exponential) with $\tau_{\omega}=11.3$~ps and $\tau_{\sigma}=11.8$~ps. The excitation temperature is $T_{\rm exc}=450$~K (see \autoref{fig:vdos_time}, \autoref{fig:300_spectral}, and \autoref{fig:stress} in section \ref{sec:NVE} of Supplementary  for data corresponding to $T_{\rm exc}=600$~K}).}
\label{fig:MD}
\end{center}
\end{figure} 

We also compute the time evolution of the frequency shift, which shows a single time constant of $\tau_{\omega} = 11.3\pm0.4$~ps for an excitation of \qty{450}{K} (\autoref{fig:MD}c, right y-axis). The observed dynamics is the combination of the phonon thermalization, \textit{i.e.} the decay of optical modes into acoustic modes, occurring within a few ps, and the following thermal diffusion. The atoms in the local area excited by the pump pulse decay to a final ground state which is hotter than the equilibrium one~\cite{DettoriJPCL19}. The resulting temperature gradient will drive the spatial diffusion (see temperature difference between excited and unexcited regions in Supplementary \autoref{fig:kinetic_temp}). The temperature difference decays with a double exponential as well, with a fast relaxation of \qty{2.96(0.03)}{ps} and a slower time constant of \qty{11.52(0.03)}{ps} for $T_{\rm exc}=450$~K, which confirms what is observed for the peak intensities in \autoref{fig:MD}b. These slower components, including the tail of the intensities and the frequency, can be traced back to the contribution of the stress induced by the {\it pump} signal and the gradual adjustment of the lattice spacing. Initially, the rapid anharmonic decay redistributes energy among modes, but the slower adjustment of the average bond distances (akin to thermal expansion) lags, leading to a more gradual frequency shift. Indeed, we calculated the atomic stress difference between the excited and unexcited regions (\autoref{fig:MD}c, left y-axis), finding rise times of \qty{11.8(0.2)}{ps} (\qty{11.7(0.3)}{ps}), which is consistent with $\tau_{\omega}$ and with the longer component of the intensity decay, $\tau_{I2}$. These results suggest that the slower behavior observed for the Raman frequency shift could also depend on the local thermal-strain relaxation: in fact, the optical phonon frequency shifts with volume (or equivalently, stress) according to its mode Gr\"uneisen parameter $\gamma$: $\Delta\omega/\omega_0=-\gamma\Delta V/V$. The excess energy delivered by the pump pulse will result in a locally expanded region, causing a red shift of the excited optical mode. Finally, we also extracted the FWHM temporal evolution of the excited peak from our MD simulations (see \autoref{fig:300_spectral}e~and~f of the Supplementary), and found it to exhibit an intermediate dynamical profile that qualitatively matches our TRRS measurements. We argue that this behavior arises from two competing processes. Following excitation, the system is highly anharmonic, resulting in a large FWHM. As phonon–phonon scattering drives thermalization of the excited mode, this anharmonicity rapidly diminishes, yielding a fast reduction in linewidth. At the same time, energy transfer from the pumped mode into other vibrational modes heats the local lattice environment. This higher local temperature sustains an enhanced scattering rate, which in turn continues to broaden the Raman peak. Thus, although the intrinsic anharmonic broadening decays quickly, the residual thermal bath remains ‘hot’ on longer timescales, maintaining the FWHM at an intermediate value until the excess heat diffuses away. Details of the computational protocol, the temperature, intensity, and stress time evolution, and convergence tests on the computational cell size can be found in \ref{sec:MD} of the Supplementary.

\section{Discussion}
TRRS and TR access the energy flow after an ultrafast excitation in $(111)$-oriented bulk Ge. We observe an increase in the TO/LO phonon temperature, which is given by the transferring of energy from photoexcited holes following ultrafast excitation, as proved by DFT calculations. We calculate a subsequent temperature decay time of \qty{2.7}{ps}, which we found to be in excellent agreement with both DFT and MD simulations, indicating an anharmonic TO/LO phonon decay into acoustic phonons as the main energy dissipation mechanism. Remarkably, the temperature decay time is strikingly different from that of the change in Raman frequency and the change in linewidth, which we attribute to time-evolving thermal strain, as proved by MD simulations. In addition, we observe Brillouin oscillations from our TR data performed under the same conditions as TRRS. This reflectivity modulation is generated from the interaction of the probe beam and a longitudinal strain pulse launched by the pump laser, which subsequently travels through Ge. Interestingly, the decoherence time of these oscillations, \qty{18}{ps}, nicely matches the decay time of the Raman linewidth. This is indicative of a correlation between the coherent acoustic phonons launched by the ultrafast laser and the TO/LO phonon mode dynamics. These results offer a unique understanding of the energy dissipation at ultrafast time scales and the entanglement between fundamental excitations and pump-induced strain, critical for the improvement of micro- to nanoelectronic devices.


\section{Methods}\label{sec4}
\renewcommand{\thefigure}{S\arabic{figure}}
\setcounter{figure}{0}
\subsection{Ultrafast spectroscopy}
TR and TRRS, are performed at room temperature on $(111)$-oriented intrinsic crystalline \ce{Ge}, which is excited with a pump pulse with a duration of \qty{30}{fs} and a wavelength of \qty{800}{nm} (\qty{1.55}{eV}). The response of the sample is monitored using a less intense probe pulse with a duration of \qty{1.5}{ps} and a wavelength of \qty{640}{nm}, which is focused on the sample with a 20x objective that provides a spot size of \qty{3}{\mu m}, while the pump size is at least twice the size of the probe. The probe fluence is set at $\leq$ \qty{0.3}{mJ/cm^{2}} while the pump fluence is varied in a range of \qtyrange{2}{9}{mJ/cm^{2}}. The pump pulse goes through a mechanical delay line, allowing for a controlled delay time between the pump and the probe within a temporal window of \qty{200}{ps}. The pump and probe pulses are in a cross-polarization configuration. We select a polarization parallel to the probe polarization in the detected signal, making it easier to filter out the pump light. Moreover, a shortpass spectral filter with a cutoff frequency of \qty{700}{nm} is also used to filter any remaining pump intensity. The probe polarization is parallel to the $(010)$ crystal direction. According to the Raman selection rules \cite{cardona_light_1975}, this orientation allows us to probe 2/3 of the TO and 1/3 of the LO phonon modes, which are degenerated at the zone center, and to which we refer as TO/LO phonon mode. For each delay time, we collect and average four spectra with a triple-spectrometer with a \qty{300}{s} integration time. For TR measurements, changes in the sample reflectivity are monitored as a function of time delay using a balanced photodiode combined with a lock-in amplifier detection scheme.

\subsection{Coupled hot hole and phonon dynamics simulations}
The hot hole dynamics is described by propagating in time the coupled time-dependent Boltzmann transport equations for carriers and phonons, in the framework of excess-energy dependent model which was described in \cite{Sjakste:2025} (see Supplementary \ref{DFT} for details). This method includes carrier-phonon and phonon-phonon collision terms, and a single effective LO/TO phonon mode. In the case of thermalised distributions which we consider in this work \cite{othonos_correlation_1991}, this approach is equivalent to  two-temperature model \cite{Caruso:2022}. Excess-energy dependent carrier-phonon collision term was modeled based on DFT calculations (see below), while phonon-phonon collision term was described by a constant decay rate based on DFT result. 

Coming to carrier distribution function, energy and band structure considerations for the single-photon absorption of \qty{1.55}{eV} photons in Ge yield holes temperature of around \qty{1500}{K}\cite{othonos_correlation_1991}. However, according to \cite{Carroll:2012}, intraband absorption  plays an important  role for holes in Ge, provided that  photoexcited carrier densities are relatively high in our experiment. While having negligible effect  on our estimation of the photoexited carrier densities, intraband absorption can lead to important changes in photoexcited hole temperatures.
Therefore, in order to account for a few percent of holes which might be concerned by intraband absorption of \qty{1.55}{eV} photons, we have considered the photoexcited hole temperature of \qty{2500}{K} for our calculation of hole-phonon energy transfer which leads to temperature rise of LO/TO phonons. 

Finally, coming to {\it ab initio} calculations, Ge was described using DFT within the LDA approximation, with lattice parameter of 10.696~a.u.\cite{Tyuterev:2011}. The rate of energy transfer from carriers to phonons was calculated by DFT-based approach  using \textsc{Quantum ESPRESSO} \cite{Giannozzi2017}, Wannier90\cite{Pizzi:2020} and EPW\cite{Ponce:2016} codes (see Supplementary \ref{DFT} for details). The rate of energy transfer from carriers to phonons changes with excess energy as the density of final states. Because of this, the calculated {\it ab initio} data could be successfully modeled with energy-dependent model of the form $C_{el-ph}(1-f(\epsilon))\sqrt{\epsilon}$, where $\epsilon$ is excess energy, $f(\epsilon)$ is hole distribution and $C_{el-ph}$ is a constant (see Supplementary \ref{DFT}). For the calculation of LO/TO phonon decay via phonon-phonon interaction, we used D3Q code~\cite{Paulatto:2013} to obtain the three-phonon anharmonic coupling rate for zone-center optical phonons.

\subsection{Molecular Dynamics simulations}
Pump-probe molecular dynamics simulations were performed with periodic boundary conditions on a bulk germanium box with length $L=85.1$~nm and section of $2.8\times2.8$~nm$^2$, for a total of 30000 atoms. Germanium atoms were modeled according to Tersoff potential\cite{Tersoff1988}. The system was initially prepared at \qty{300}{K} with an NPT run by means of the Nos\'e-Hoover thermostat for \qty{50}{ps} with a timestep of \qty{1}{fs}, in order to reach the equilibrium density at \qty{300}{K}. Then, a NVT run was performed with the stochastic velocity rescaling thermostat\cite{Bussi2007} for further \qty{50}{ps} with a timestep of \qty{1}{fs}. 

From this initial setup, the system is divided in three distinct regions to model the experimental conditions of a laser hitting a localized spot in the sample. As shown in the schematic representation of \autoref{fig:GLE_set}a, the central area which is $2.8$~nm-long, is excited according to the Generalized Langevin Equation (GLE) thermostat while the surrounding regions are left unperturbed at room temperature. The GLE only acts on the system for \qty{500}{fs}, realizing an initial excitation of $T_{\rm exc}=450$~K or \qty{600}{K}, corresponding to the temperature rise observed in the experiments. After the excitation, the thermostat is turned off and the simulation cell evolves freely in the NVE ensemble towards its final equilibrium configuration. The corresponding temperature profile is reported in \autoref{fig:GLE_set}b~and~c for $T_{\rm exc}=450$~K and \qty{600}{K} respectively: a $t=0$, the system is characterized by a hot-spot which progressively smooths out during the relaxation. The results reported in this work are obtained by performing a configurational average over 100 statistically independent runs (\textit{ie} initialized with different initial atomic velocities).

\backmatter
\clearpage

\bmhead{Supplementary information}
The Supporting Information is available free of charge at \textit{(to be added upon acceptance)}.

\bmhead{Data Availability}
All data within the article and the Supplementary Information that support the findings of this study are openly available in ZENODO at \textit{(to be added upon acceptance)}, Reference No. \textit{(to be added upon acceptance)}.




\bmhead{Author contributions}
I.Z. conceived the experiment and supervised the project. G.R., A.K.S., J.M.S.-G. and B.A. developed the setup and measurements protocol. G.R. and B.A. performed the time-resolved Raman and transient reflectivity measurements and analyzed the data. J.S., R.S., and N.V. performed the DFT calculations. R.D., C.M., and R.R. performed the MD simulations. All authors discussed the results and either wrote or reviewed the paper.

\bmhead{Acknowledgements}

I.Z. gratefully acknowledges support from the Swiss National Science Foundation grant (Grant No. \textbf{189924} and Grant No. \textbf{170741}) and from the European Research Council (ERC) under the European Union's Horizon 2020 research and innovation program (Grant No \textbf{756365}). B.A. acknowledges support from the European Union’s Horizon 2020 research and innovation programme under the Marie Skłodowska-Curie (Grant No. \textbf{891443}), as well as the Swiss National Science Foundation PRIMA grant (Grant No. \textbf{208608}). J.S., R.S. and N.V. acknowledge financial support from the ANR DragHunt project ANR-24-CE50-3505. Computer time has been granted by the national centers GENCI-CINES and GENCI-TGCC (Project 2210), and by \'Ecole Polytechnique through the 3Lab computing cluster.
\begin{appendices}




\end{appendices}


\bibliography{sn-bibliography}

\clearpage


\input{0_SUPPLEMENTARY}

\end{document}

%% file: 0_SUPPLEMENTARY.tex
\renewcommand{\thesection}{S\arabic{section}}
\section*{\LARGE Supplementary information: \newline
Unraveling energy flow mechanisms in semiconductors by ultrafast spectroscopy: Germanium as a case study}
\setcounter{section}{1}


Grazia Raciti$^1$, Bego\~{n}a Abad$^1$, Riccardo Dettori$^2$, Raja Sen$^3$, Aswathi K. Sivan$^1$, Jose M. Sojo-Gordillo$^1$, Nathalie Vast$^4$, Riccardo Rurali$^5$, Claudio Melis$^2$, Jelena Sjakste$^4$, and Ilaria Zardo$^1$

\begin{enumerate}
\item{Department of Physics, University of Basel, 4056 Basel, Switzerland}

\item{Department of Physics, University of Cagliari, CA 09042 Monserrato, Italy}

\item{SATIE, CNRS, ENS Paris-Saclay}, \orgname{Université Paris-Saclay, 91190 Gif-sur-Yvette, France}

\item{Laboratoire des Solides Irradi\'es, CEA/DRF/IRAMIS, Ecole Polytechnique, CNRS, Institut Polytechnique de Paris, 91128 Palaiseau, France}

\item{Institut de Ci\`encia de Materials de Barcelona, ICMAB--CSIC, Campus UAB, Bellaterra, 08193, Spain}
\end{enumerate}
\clearpage
\subsection{Experimental details} \label{Exp_details}
\subsubsection{Multi-scheme pump-probe setup}

\begin{figure}[b]
\begin{center}
\includegraphics[width=0.8\textwidth]{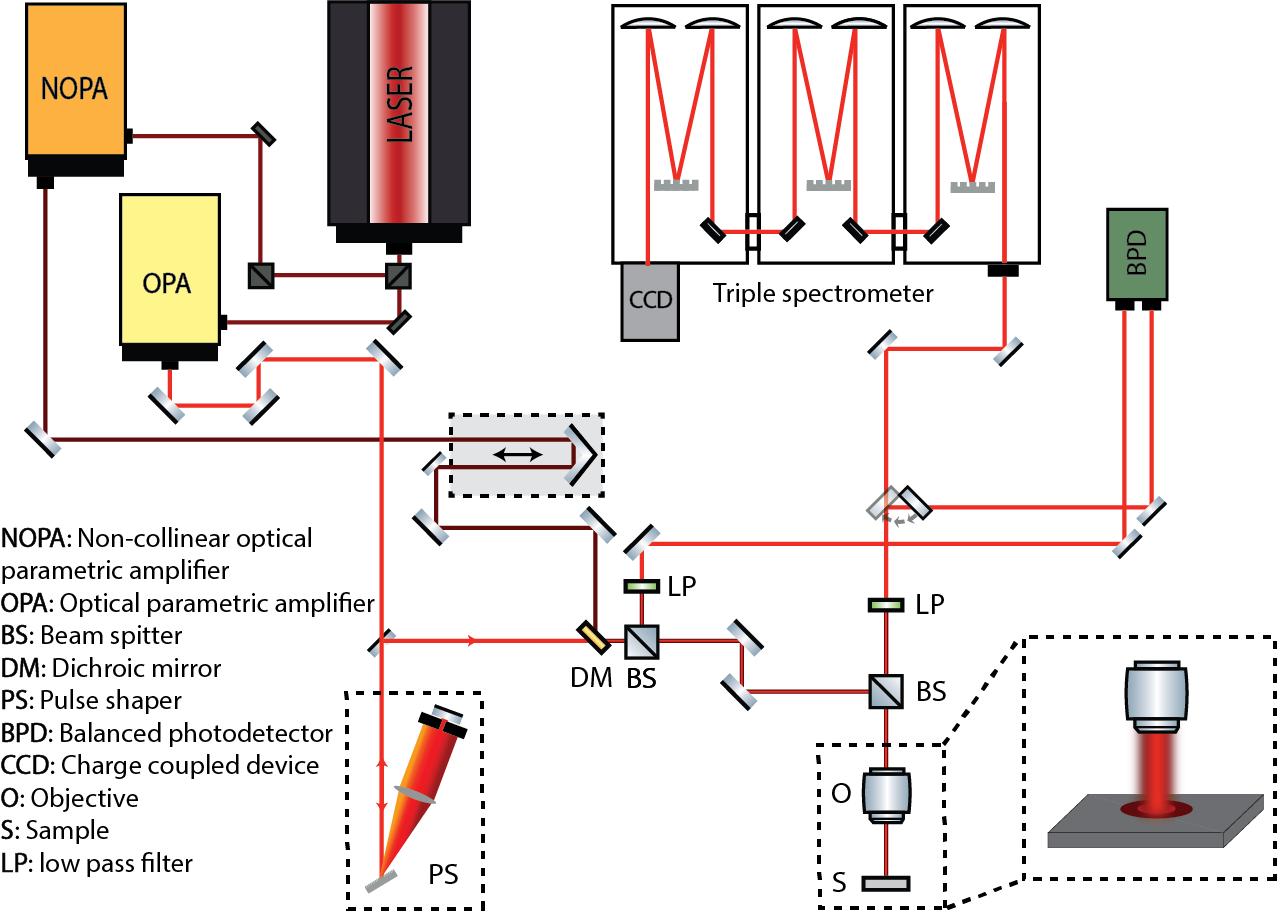}
\caption{\textbf{Time-resolved Raman and transient reflectivity pump-probe setup}. A femtosecond pulse laser (\qty{1040}{nm}) is split into two beams that are steered to optical parametric amplifiers and that will act as a pump and probe. A pulse shaper (PS) is used to stretch the pulse up to \qty{1.5}{ps} in order to increase the spectral resolution (\qty{14}{cm^{-1}}) for Raman spectroscopy measurements. Both beams are shined on the sample by using a 20x objective. The backscattered light is then collected and analyzed either through a triple spectrometer (TRRS) or a balanced photodiode (TR).}
\label{fig: Schematics}
\end{center}
\end{figure} 

\autoref{fig: Schematics} shows a scheme of the experimental setup used to perform time-resolved Raman spectroscopy (TRRS) or transient reflectivity (TR) measurements, depending on the detection scheme. Both techniques rely on an ultrafast hybrid ytterbium fiber laser from Spectra-Physics (Spirit 1030-70) with a wavelength of \qty{1030}{nm}, a pulse duration of hundreds of femtoseconds ($<400$ fs), and a repetition rate of \qty{1}{MHz}. The output of this laser is split into pump and probe pulses, which are steered to a non-collinear optical parametric amplifier (NOPA) and an optical parametric amplifier (OPA), respectively, that can tune the wavelength of the ultrafast beams in a range of \qtyrange{600}{900} {nm}.\\
In the case of TRRS measurements, Raman spectra are tracked as a function of delay time between pump and probe. Remarkably, the OPA output beam has a pulse duration of \qty{180}{fs} and goes directly to a pulse shaper, which stretches the temporal duration by one order of magnitude, up to \qty{1.5}{ps}. This is crucial to improving the spectral resolution of the Raman measurements as time and frequency are Fourier conjugates, and the shorter the pulse, the larger its spectral bandwidth. Stretching the pulse enables us to achieve a spectral resolution of \qty{12.5}{cm^{-1}}, with which we can resolve the different spectral features of the probed Raman spectra. This probe is used to measure Raman spectra at different times before and after pump excitation, acquired by a triple spectrometer (TriVista TR 555) with three \qty{1500}{l/mm}, \qty{1500}{l/mm}, and \qty{1800}{l/mm} gratings. The dispersed light is collected with a Princeton Instrument charge-coupled device (CCD) (ProEM+). \\
For TR measurements, the change in reflectivity of the sample surface is tracked as a function of the delay time between the pump and probe pulses. TR uses modulation of the pump laser by a chopper whose frequency is locked to 666 Hz and is used as a reference for a lock-in amplifier (SR830). This approach enables filtering of the noise at various frequencies and amplifying the signal that is detected by a balanced photodiode (Physics Basel, SP 1’023). \\
While the selection of the pump-probe temporal overlap ($t_0$) in standalone TR experiments has little transcendence, in this work, $t_0$ is consistently used as the temporal reference for our TRRS experiment as well. This ensures a perfect synchronization between the TR signal and the dynamics observed by TRRS. Here, $t_0$ is defined as the time in which the TR signal starts to rise, since the reflectivity response immediately changes with the creation of charges by the pump pulse excitation, its temporal resolution is given by the cross-correlation in between the pump and probe pulses, whose durations are \qty{30}{fs} and \qty{1.5}{ps}, respectively.

\subsubsection{TRRS signal processing} \label{TRRS_processing}
\begin{figure}[b]
    \centering
    \includegraphics[width=\linewidth]{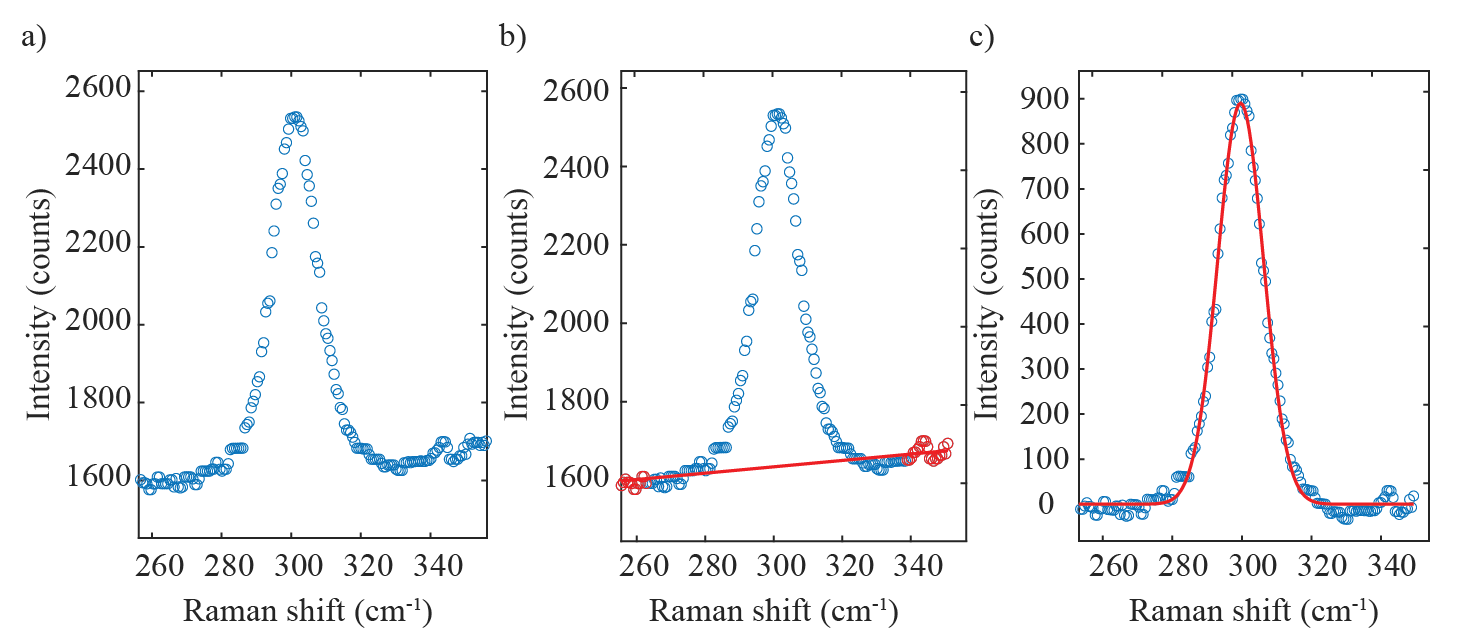}
    \caption{a)~Representative raw Ge Raman spectrum. b)~Linear fit of the baseline to remove the background signal. The red data points are the ones selected to perform the linear fit represented by the solid red line. c)~Raman spectrum after the baseline subtraction. The red solid line represents the Gaussian fit.}
\label{fig:DataAnylisis}
\end{figure}
The collection of the data is automated with a self-written Python script. Starting from the anti-Stokes side, a spectrum per time-step of the delay line is collected. Before moving the spectrometer grating to the Stokes side, we acquire a neon lamp spectrum, which, as explained afterwards, serves to calibrate the spectral window. We then proceed with the acquisition of the Stokes spectra using the same conditions as for the AS. After this, the neon lamp spectrum is acquired again to calibrate the Stokes spectral window. When the measurement is completed, two sets of data, one for S and the other for AS, as a function of the delay time between the pump and the probe, are generated. Absolute frequency calibration of the Raman spectra was done by using the well-known Neon lamp emission lines, recorded at the end of each dataset, as reference points to correct the spectral axis. All the data analysis is automated with a self-written MATLAB script. In the following, we describe the data analysis steps for a representative spectrum; the same procedure was then applied to the entire dataset. A representative raw Ge Raman spectrum is plotted in \autoref{fig:DataAnylisis}a. It exhibits a peak at $\approx$ \qty{300}{cm^{-1}}, which is the transverse optical (TO)/longitudinal optical (LO) degenerate phonon mode in bulk Ge. First, we use a linear fit to account for the background signal in the spectrum. The red points in \autoref{fig:DataAnylisis}b represent the selected points used to determine the linear fit of the background, which is shown with the red solid line. This linear fit is then subtracted from the data. The resulting spectrum displayed in \autoref{fig:DataAnylisis}c is then fitted using a Gaussian function given by: 
\begin{equation}
   y_{G}= y_0 + \frac{A}{w\sqrt{\pi/2}} e^{ \frac{-2(x - x_0)^2}{w^2}}
\end{equation}
where $y_0$, $x_0$, $A$, $w$ represents offset, center, linewidth and area respectively. We use a Gaussian function because the Gaussian spectral lineshape of our probe pulse dominates over the Lorentzian lineshape of the phonon mode, as the FWHM at room temperature is \qty{13}{cm^{-1}} for the probe pulse and \qty{2.5}{cm^{-1}} for the phonon mode~\cite{Menendez_1984, Tang_1991}. The intensity of the peak is calculated from the fit parameters as follows:
\begin{equation}
I = y_0 + \frac{A}{w \sqrt{\pi/2}}.
\end{equation}
In addition, in \autoref{fig:Intensity} we show the difference spectra, which are calculated by subtracting the average intensity of all spectra acquired before excitation from each spectrum at selected times. Moreover, to avoid artifacts in the resulting difference, the dynamic Raman frequency change, $\Delta \omega(t)$, must be taken into account and suppressed. For this purpose, we extract the average Raman shift of all spectra acquired before excitation ($\overline{\omega}(t<0)=\pm$ \qty{301}{cm^{-1}}) and subtracted it from the Raman shift at each selected time, $\omega(t)$, obtaining $\Delta \omega(t)$. The difference between these values, $\omega(t)-\Delta \omega(t)$, is the corrected Raman frequency, $\omega_{corr}(t)$, which is extracted from each spectrum displayed in \autoref{fig:Intensity}c~and~d.

\subsubsection{Uncertainty calculation}
The shading in \autoref{fig:PhonProp} corresponds to the experimental error, which is calculated from the standard deviation of each property (fitted intensity, change in Raman frequency, and change in linewidth) before any pump excitation, i.e., before \qty{0}{ps}. The error in the change in frequency and linewidth of the S band is smaller than the AS one, since the S signal is higher because of its larger probability. In contrast, the fitted intensity error is greater for S scattering, since $\Delta I/I$ is larger for AS scattering, as shown in \autoref{fig:Intensity}.

\subsubsection{Experimental phonon temperature calculation} \label{Ph_temp}
The intensities of the S and AS bands are related to the phonon population, from which the temperature of the phonon mode can be calculated \cite{prasankumar_optical_2012}. Indeed, the temperature of the phonon mode can be tracked by measuring the change of any of the spectral properties, Raman frequency, linewidth, or ratio of S and AS intensities. This approach is the well-known Raman thermometry that has been used to extract temperature maps and thermal properties of numerous materials \cite{sandell_thermoreflectance_2020}. This technique is traditionally performed under equilibrium conditions using continuous-wave laser sources to probe. While conventional Raman thermometry probes the average temperature of the lattice, ultrafast Raman thermometry measures the temperature of the probed optical phonon mode at a specific time after excitation \cite{prasankumar_optical_2012}. Importantly, Raman thermometry relies on a prior calibration, in which the sample is heated up to correlate phonon spectral features such as linewidth and frequency with temperature. This calibration is essential to extract accurate temperature values from spectral changes. However, an alternative approach allows one to extract the temperature from the ratio between the AS and S intensity, as \autoref{eq:AS_over_S} indicates \cite{olsson_magnon_2018}, without any prior calibration of the Raman properties as a function of temperature: 
\begin{equation}
    \frac{I_{\text{AS}}}{I_{\text{S}}} = C_{\text{exp}} \left(\frac{\omega_{\text{exc}} + \omega}{\omega_{\text{exc}} - \omega}\right)^4 e^{-\frac{\hbar \omega}{k_B T}}
    \label{eq:AS_over_S}
\end{equation}
where $C_\text{exc}$ is an empirically-determined constant related to the experimental differences in the light collection between S and AS spectral windows given by the different instrument response at these relatively large separations of frequencies. In this equation, $\omega$ is the frequency of the probed phonon, $T$ is the temperature, and $\omega_\text{exc}$ is the excitation frequency. Therefore, we extract the temperature from this intensity ratio after determining $C_\text{exp}$ by calibrating our experimental setup, as explained in the next section.

\subsubsection{Temperature calibration}
\begin{figure}
\begin{center}
\includegraphics[width=\textwidth]{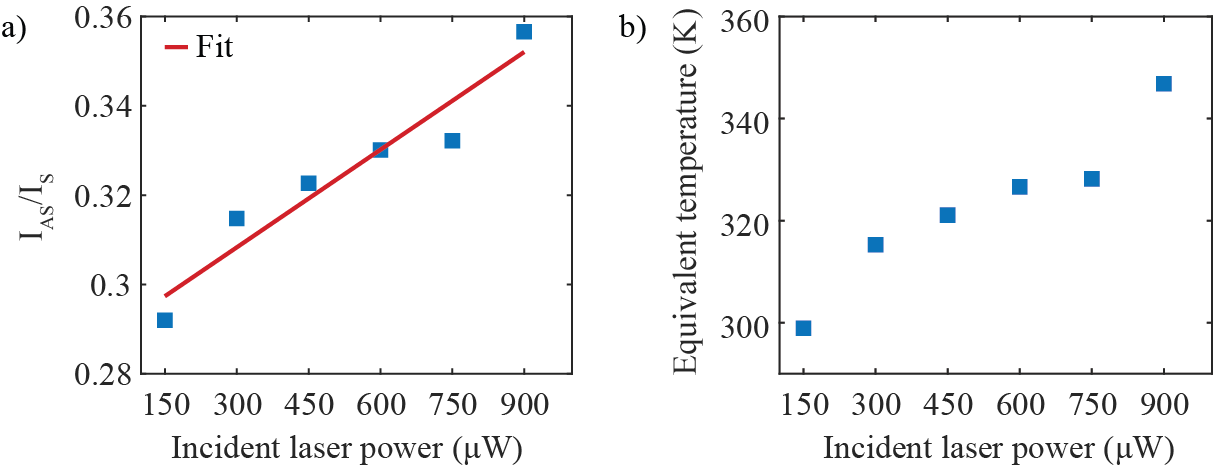}
\caption{a)~Intensity ratio between AS and S signals at each probe laser power (blue squares). The red solid line represents the linear fit used to extract the value of the intensity ratio to the zero heating power, which is needed to determine the calibration factor. b)~Equivalent temperature calculated by using the calibration factor and employed to determine the probe power to perform the experiments, avoiding heating effects.}
\label{fig:Cal_Ge}
\end{center}
\end{figure}

To determine $C_{exp}$, we perform a laser probe power study at room temperature (\qty{300}{K}). We perform steady-state Raman spectroscopy using only the probe pulse. We collect different Raman spectra for both S and AS, varying the incident probe laser power from \qtyrange{150}{900}{\mu W}. Then, for each laser power, the corresponding intensity ratio is calculated as shown in \autoref{fig:Cal_Ge}a. 
We extrapolate the intensity ratio to zero laser heating by using a linear fit function (red solid line in \autoref{fig:Cal_Ge}a). Plugging this value in \autoref{eq:AS_over_S} we extract a calibration factor of 1.04. We can use this calibration factor to determine the equivalent temperature of the phonon mode at different probe powers. The results are shown in \autoref{fig:Cal_Ge}b. The increase in temperature with incident laser power indicates that we are heating the sample. In addition, this procedure is useful to determine the power level used in the experiment that minimizes the probe heating of the sample, which could affect the measured dynamics. Therefore, we found a value of \qty{150}{\mu W} for the probe laser to be a good compromise between obtained signal-to-noise ratio and the resulting sample heating.

\subsubsection{Brillouin oscillations} \label{sec:Brillouin_osc}

We observe low-frequency oscillations ($\sim$ tens of GHz) from our transient reflectivity signal which we identify as Brillouin oscillations \cite{ezzahri_coherent_2007}. They are detected from the interference between the probe light reflected by the surface and the probe light reflected from the moving strain wave \cite{wright_thickness_1992}. Indeed, the frequency of the Brillouin oscillations, in the case of normal incidence, is given by:
\begin{equation}
    f_{Brillouin} = \frac{2 v_Ln(\lambda)}{\lambda}
    \label{eq:brillouin}
\end{equation}
where $v_L$ and $n(\lambda)$ are the longitudinal-acoustic sound velocity and the refractive index at the probe wavelength $\lambda$. The longitudinal acoustic sound velocity for \ce{Ge} is approximately $v_L = 5400$ m/s \cite{aspnes_dielectric_1983}, while the refractive index is 5.4067 \cite{aspnes_dielectric_1983}. The expected Brillouin oscillation at the experimental wavelength of the probe \qty{640}{nm} has a frequency of $f_{Brillouin} = 91.13$ GHz, in excellent agreement with what we measured.

\subsection{Ab initio calculations of electron-phonon and phonon-phonon interactions} \label{DFT}
\subsubsection{Computational details}
In this work, germanium is described within density functional theory (DFT) and density functional perturbation theory (DFPT) using \textsc{Quantum ESPRESSO}, Wannier90, EPW and D3Q  codes. 
Germanium was described within the LDA approximation, with a cutoff energy of 60 Ry and a 12x12x12 $\Gamma$-centered $\mathbf{k}$-point grid. The lattice parameter of 10.696~a.u.\cite{Tyuterev:2011} was used in all the calculations
with the \textsc{Quantum ESPRESSO} package~\cite{Giannozzi2017}. 

The Wannierization parameters used for the Wannier90 code~\cite{Pizzi:2020} were as follows: 8 Wannier functions were constructed from 10 initial Bloch states, using atom-centered sp$^3$  projections as the initial guess. A disentanglement window of 17 eV and a frozen window of 12.203 eV were applied, along with a 16 × 16 × 16 $\Gamma$-centered $\mathbf{k}$-point grid. Disentanglement convergence was achieved within 109 iterations, with a tolerance of $10^{-12}$ \AA$^2$, while the spread minimization reached a tolerance of $10^{-10}$ \AA$^2$ after 200 iterations. The total spread for all eight Wannier functions was 33.45 \AA$^2$. For the interpolation of the electron-phonon matrix elements within the EPW code~\cite{Ponce:2016}, we used the above parameters and a 8x8x8 $\mathbf{q}$-point grid.

For the calculation of LO/TO phonon decay via phonon-phonon interaction, we used D3Q code~\cite{Paulatto:2013}, under a very fine internal grid of 39 × 39 × 39 q-points and a Gaussian smearing of 2 cm$^{-1}$.

\subsubsection{Energy transfer from carriers to phonons }\label{sec_DFT_energy_transfer}

The rate of energy transfer from a single carrier to phonons reads \cite{Allen:1987}:

\begin{equation}
\label{en_relax}
\begin{split}
\frac{\delta E}{\delta t}_{n\mathbf{k}}=\\ 
2\pi  \sum_{m\nu}\int\frac{d\mathbf{q}}{\Omega_{BZ}}\omega_{\mathbf{q}\nu}|g_{mn\nu}(\mathbf{k},\mathbf{q})|^2
(1-f_{m,\mathbf{k}+\mathbf{q}})(N_{\mathbf{q},\nu}+1)
\delta(\varepsilon_{n,\mathbf{k}}-\varepsilon_{m,\mathbf{k}+\mathbf{q}}-\hbar\omega_{\mathbf{q}\nu})-\\
-2\pi\sum_{m\nu}\int\frac{d\mathbf{q}}{\Omega_{BZ}}\omega_{\mathbf{q}\nu}|g_{mn\nu}(\mathbf{k},\mathbf{q})|^2
(1-f_{m,\mathbf{k}+\mathbf{q}})N_{\mathbf{q},\nu}\delta(\varepsilon_{n,\mathbf{k}}-\varepsilon_{m,\mathbf{k}+\mathbf{q}}+\hbar\omega_{\mathbf{q}\nu}) 
\end{split}
\end{equation}

Here, 
$g_{mn\nu}(\mathbf{k},\mathbf{q})$ is the electron-phonon matrix element, which depends on the initial electronic state $|n,\mathbf{k}\rangle$ with band number $n$ and wavevector $\mathbf{k}$, on the phonon $|\nu,\mathbf{q}\rangle$, where $\nu$ is phonon mode number and $\mathbf{q}$ phonon wave vector, and on the final electronic state $|m,\mathbf{k+q}\rangle$.
The Dirac delta functions
$\delta(\varepsilon_{n,\mathbf{k}}-\varepsilon_{m,\mathbf{k+q}}
-\hbar\omega_{\mathbf{q}\nu})$ and $\delta(\varepsilon_{n,\mathbf{k}}-\varepsilon_{m,\mathbf{k+q}}
+\hbar\omega_{\mathbf{q}\nu})$ 
represent the energy conservation laws for respectively phonon emission and absorption. $f_{n,\mathbf{k}}$ is the carrier distribution function. Note that the phonon occupations $N_{\mathbf{q},\nu}$ depend on lattice temperature $T_L$, while the carrier Fermi-Dirac distribution function depends on carrier temperature $T_c$ and on the chemical potential $\mu$. 

The total rate of energy transfer from carriers to phonons for a given initial carrier distribution reads:
\begin{equation}
\frac{\delta E}{\delta t}_{c-ph}= 
\sum_{n}\int\frac{d\mathbf{k}}{\Omega_{BZ}}
\frac{\delta E}{\delta t}_{n\mathbf{k}} f_{n,\mathbf{k}} 
\end{equation}

\begin{figure}[ht]
\begin{center}
\includegraphics[width=\textwidth]{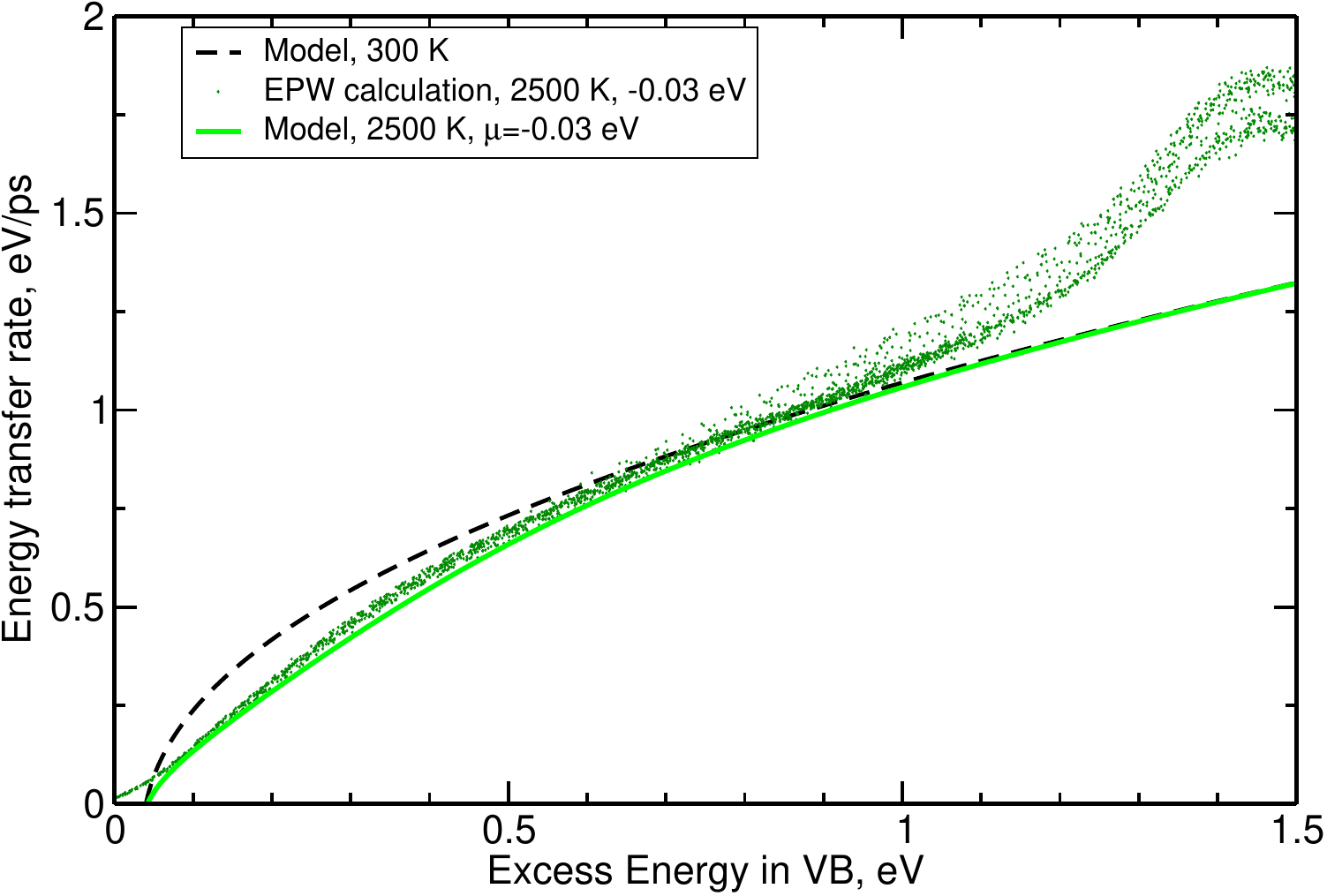}
\end{center}
\caption{The rate of energy transfer $\frac{\delta E}{\delta t}_{n\mathbf{k}}$ from photoexcited holes close to valence band maximum to TO/LO phonons  calculated using modified EPW code, and compared to the excess-energy dependent model $\frac{\delta E}{\delta t}(\epsilon)$ (see text).}
\label{fig:el-ph-rate}
\end{figure}

The rate of energy transfer $\frac{\delta E}{\delta t}_{n\mathbf{k}}$ from photoexcited holes close to valence band maximum to TO/LO phonons  was calculated using modified EPW code on 30x30x30 $\mathbf{q}$-grid (convergence checked with 50x50x50 grid). An example of such calculation is shown in \autoref{fig:el-ph-rate}, where calculated $\frac{\delta E}{\delta t}_{n\mathbf{k}}$ for $\mathbf{k}$-points belonging to 100x100x100 grid is shown as a function of hole excess energy $\epsilon$ for hole distribution with $T_c=2500$~K and $\mu=-0.03$ eV, which is close to the valence band maximum (shown in green dots). The rate of energy transfer from carriers to phonons changes with excess energy as the density of final states (DOS), as already reported in previous works (see e.g.~\cite{Sjakste:2025}). Because of this, the calculated {\it ab initio} data can be successfully modeled with energy-dependent model of the form $\frac{\delta E}{\delta t}(\epsilon)=C_{el-ph}(1-f(\epsilon))\sqrt{\epsilon}$, where $C_{el-ph}$ is a constant. This can be seen in \autoref{fig:el-ph-rate}: the model (clear green line) and the {\it ab initio} data are hardly distinguishable for excess energies below 1 eV. Note that the excess energies of interest in this study do not exceed \qty{0.5}{eV}. The Pauli blocking factor $1-f(\epsilon)$ plays an important role at high temperatures and low excess energies, whereas for $T_c=300$~K (black dashed line), the square-root dependence of the rate of energy transfer on the excess energy is clearly visible. Note that once the $C_{el-ph}$ is obtained from DFT calculation and validated by comparing DFT and model results, the hot hole dynamics can be described in the framework of the two-temperature excess-energy dependent model as described below. 

\subsubsection{Hot hole dynamics}\label{sec:hotholes}

The hot hole dynamics is described by propagating in time the coupled time-dependent Boltzmann transport equations (t-BTE) for carriers and phonons, in the framework of excess-energy dependent model which was described in \cite{Sjakste:2025}. In the case of the thermalised distributions which we consider in this work, this model is equivalent to two-temperature model\cite{Caruso:2022}. The detailed expressions for carrier-phonon collision term can be found e.g. in \cite{Sjakste:2025}.

The t-BTE for carriers reads:
\begin{equation}
\frac{\delta f(\epsilon)}{\delta t}=\left.\frac{\partial f(t)}{\partial t}\right\vert_{\rm{el-c}}
\label{BTE_el}
\end{equation}
Here,  $t$ is time. The collision term on the right-hand side accounts for carrier-phonon interactions. 

The t-BTE for phonons reads:
\begin{equation}
\frac{\delta N_{TO/LO}(T_L)}{\delta t}=\left.\frac{\partial N_{TO/LO}(T_L,t)}{\partial t}\right\vert_{\rm{coll}}
\label{BTE_ph}
\end{equation}
Here, the phonons are represented by one effective TO/LO phonon mode. The collision term on the right-hand side accounts for phonon-hole interactions which lead to the increase of phonon temperature and for the decay of TO/LO modes into acoustic phonons, with the decay constant calculated {\it ab initio} with the D3Q code.   

The coupled t-BTEs for carriers (holes) and phonons are solved by time-stepping with a time step of \qty{1}{fs}. 

Note that  as the holes transfer energy only to the part of the Brillouin zone (BZ) close to the zone center, the Raman-active optical phonons in this part of the BZ become overheated with respect to average temperature of optical phonons in the BZ.
To account for this, we define the "local" temperature of the Raman-active modes $T_R$:
$T_R=\frac{\Omega_{BZ}}{\Omega_{R}}T_L$.  Here, $\frac{\Omega_{BZ}}{\Omega_{R}}$
is the ratio between BZ volume and the heated part of BZ. This coefficient is determined from the calculated carrier-phonon $\mathbf{q}$-dependent spectral function for energy transfer (per carrier), which is defined as follows:

\begin{equation}
\label{en_relax_spectral}
\begin{split}
\frac{\delta E}{\delta t}(|\mathbf{q}|)= 
2\pi  \sum_{m\nu}\int\int\frac{d\mathbf{q'}}{\Omega_{BZ}}
\frac{\delta E}{\delta t}_{\mathbf{k},\mathbf{k+q'}}\delta(|\mathbf{q}|-|\mathbf{q'}|)
\end{split}
\end{equation}

\begin{figure}[!h]
\begin{center}
\includegraphics[width=\textwidth]{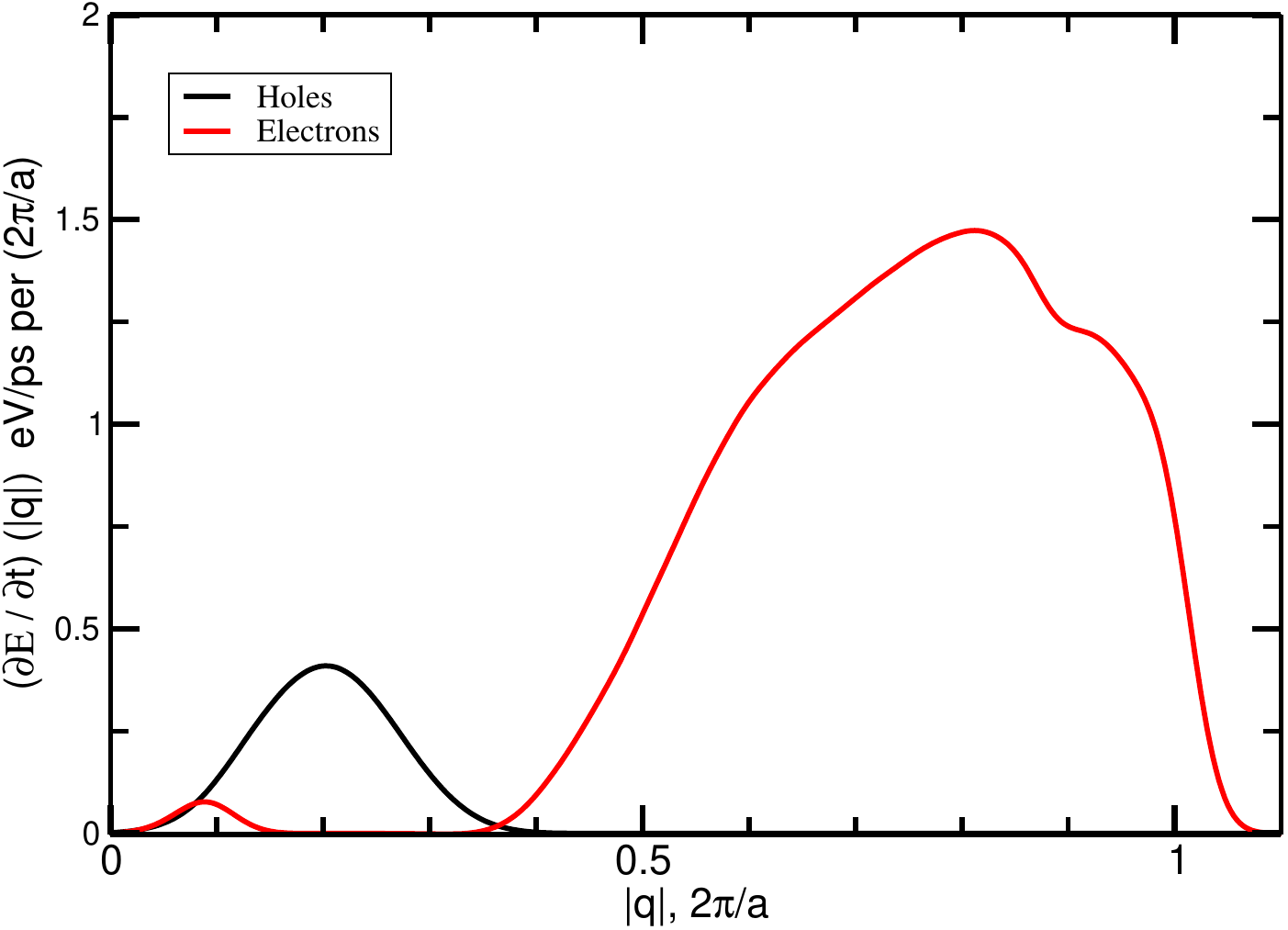}
\end{center}
\caption{The $\mathbf{q}$-dependent spectral functions for the energy transfer between carriers and optical phonons (see text), calculated per carrier for electrons and for holes in Ge.}
\label{fig:el-ph-spect}
\end{figure}

Examples of calculated $\mathbf{q}$-dependent spectral function for the energy transfer between carriers and optical phonons is shown in \autoref{fig:el-ph-spect} for holes and for electrons in Ge.  In our calculations, $\frac{\Omega_{BZ}}{\Omega_{R}}$ for holes is considered to be between 3.5 and 4, which corresponds to the heating of TO/LO phonons in 25-30\% of  BZ, as one can see in \autoref{fig:el-ph-spect}. \autoref{fig:el-ph-spect} also shows why  the heating of Raman-active modes is dominated by holes: as one can see, electrons transfer their energy to large-$\mathbf{q}$ (intervalley) phonons.

\subsection{Molecular Dynamics calculations} \label{sec:MD}
\subsubsection{Simulation setup}

\begin{figure}[!h]
\begin{center}
\includegraphics[width=\textwidth]{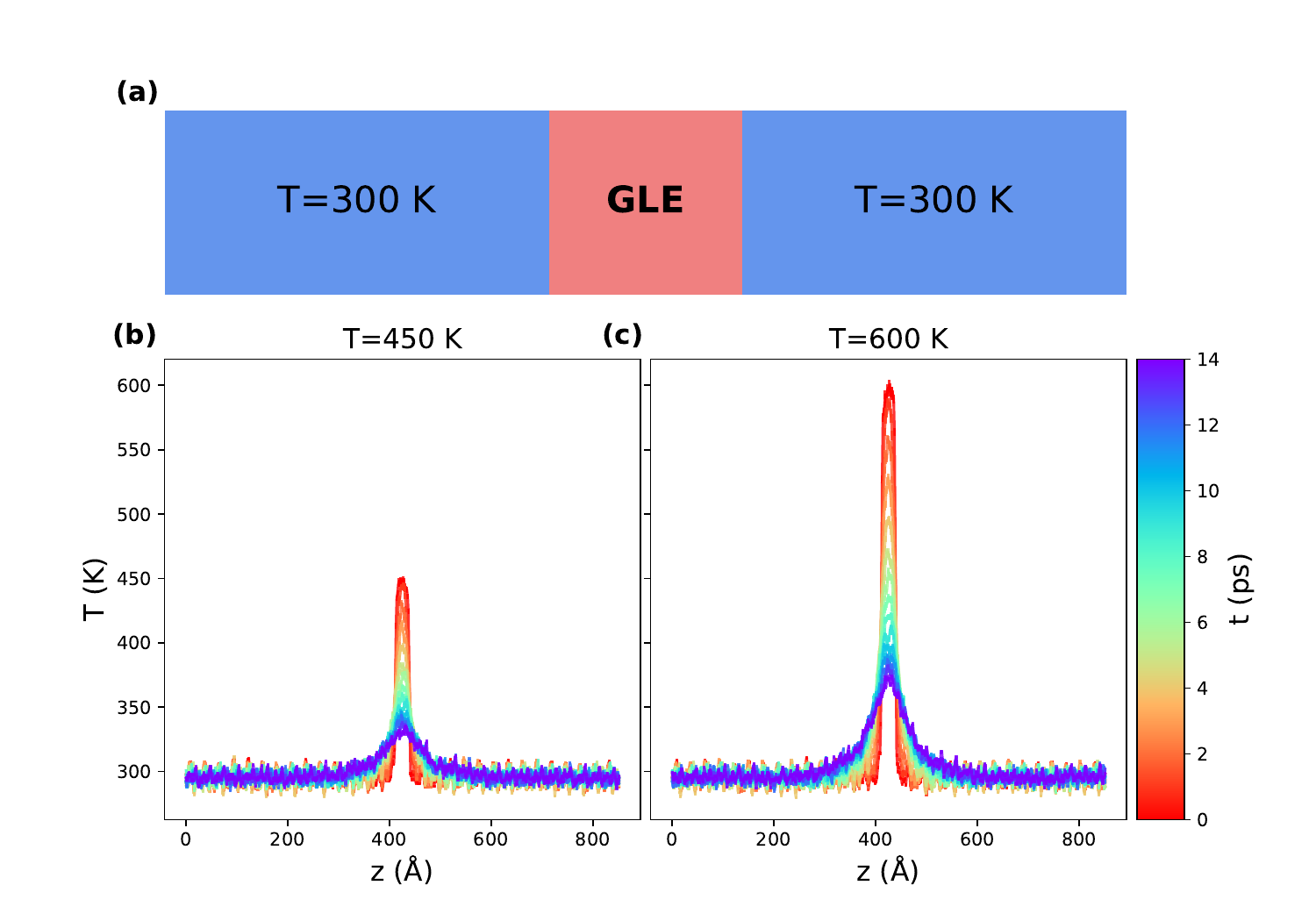}
\caption{a) Schematic representation of the simulation cell used in the MD simulations. b-c) Temperature profile as a function of time during the relaxation for the two excitation temperature considered in our simulations.}
\label{fig:GLE_set}
\end{center}
\end{figure}


\subsubsection{Generalized Langevin Equation excitation}
Atoms in the central region were excited by means of the GLE thermostat~\cite{CeriottiPCS10,CeriottiPRL09,DettoriJCTC17}. The GLE implements non-Markovian dynamics by introducing history-dependent terms~\cite{Zwanzig2001-xn} in the standard Langevin equation. In this implementation, it is designed with a standard white noise Langevin thermostat with friction $\gamma_{\rm base}$ and target temperature $T_{\rm base}$, and a $\delta-$thermostat~\cite{Ceriotti2010} at a temperature $T_{\rm exc}$ that is coupled with a friction parameter $\gamma_{\rm exc}$. The memory kernel power spectrum reads as
\begin{equation}
   K(\omega)=2\gamma_{\rm base}+\dfrac{\gamma_{\rm exc}}{\pi}\dfrac{\omega_{\rm exc}\Delta \omega \omega^2}{(\omega^2-\omega^2_{\rm exc})^2+\Delta\omega^2 \omega^2} 
\end{equation}

The friction parameters were chosen to be $\gamma_{\rm exc}=0.5$~ps$^{-1}$ and $\gamma_{\rm base}=0.5$~ps$^{-1}$. The thermostat was tuned at the optical phonon peak frequency for the Tersoff potential, $\omega_{\rm exc}=310$~cm$^{-1}$, while the frequency window was chosen as $\Delta \omega=0.1$~cm$^{-1}$ in order to minimize any energy spill‑over arising from unavoidable mode coupling, which does not guarantee a perfect adiabatic thermalization of the stretching mode. However, such a coupling does not interfere with the relaxation dynamics~\cite{DettoriJCTC17}. 

The system excitation and the relaxation is mainly monitored using the kinetic temperature, while the spectral features were extracted from a time-resolved vibrational density of states (vDOS) that is calculated every 500~fs by means of a short-time Fourier transform of the atomic velocity autocorrelation function (VACF) as
\begin{equation}
g(\omega,t)=\frac{1}{2\pi\,N}\sum_{i=1}^{N}\int_{t_1}^{t_2}
w(\tau)\,\bigl\langle \mathbf{v}_i(t)\cdot\mathbf{v}_i(t+\tau)\bigr\rangle
\,e^{-i\omega\tau}\,d\tau
\end{equation}
where the VACF is evaluated at the time $t$, $w(\tau)$ is a real–valued window function (Blackman-Nuttall~\cite{DettoriJPCL19}) of width $\Delta t=t_2-t_1$ that localizes the Fourier transform around $t$, and $N$ is the total number of atoms. The $g(\omega,t)$ is then used to calculate the intensity of the peaks and the frequency shifts as a function of time as
\begin{equation}
\label{eq:spectral_features}
\begin{split}
\mathcal{I}(\omega,t)&=\int_{\omega_1}^{\omega_2}g(\omega^\prime,t)d\omega^\prime\\
\omega(t)&=\dfrac{\int_{\omega_1}^{\omega_2}\omega^\prime g(\omega^\prime,t)d\omega^\prime}{\int_{\omega_1}^{\omega_2}g(\omega^\prime,t)d\omega^\prime}
\end{split}
\end{equation}

\subsubsection{NVE relaxation}\label{sec:NVE}
\begin{figure}[!h]
\begin{center}
\includegraphics[width=\textwidth]{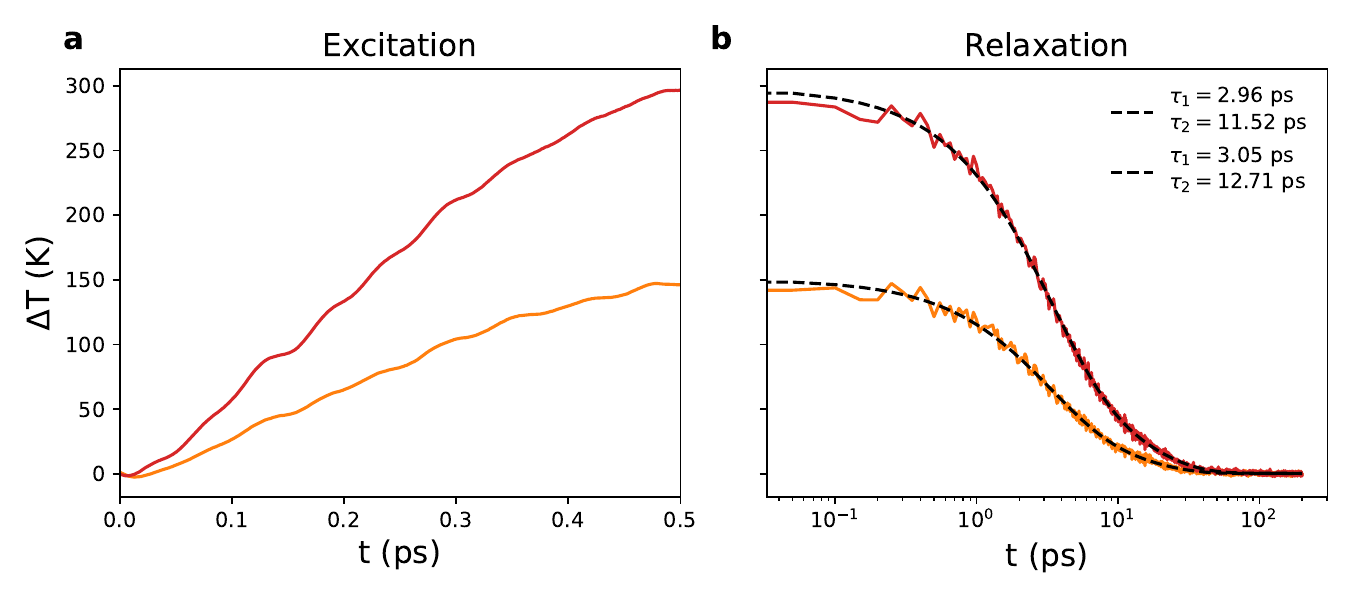}
\caption{Temperature difference between the excited and unexcited regions during (a) the 500~fs-long excitation, and (b) during the relaxation. Here temperatures are reported with a log scale along with their fit (dashed black lines) using a two exponential model.}
\label{fig:kinetic_temp}
\end{center}
\end{figure}

\paragraph{Kinetic temperature} 
During the relaxation we can monitor the system evolution by computing the kinetic temperature difference between the two regions, as shown in \autoref{fig:kinetic_temp}

\begin{figure}[!h]
\begin{center}
\includegraphics[width=0.98\textwidth]{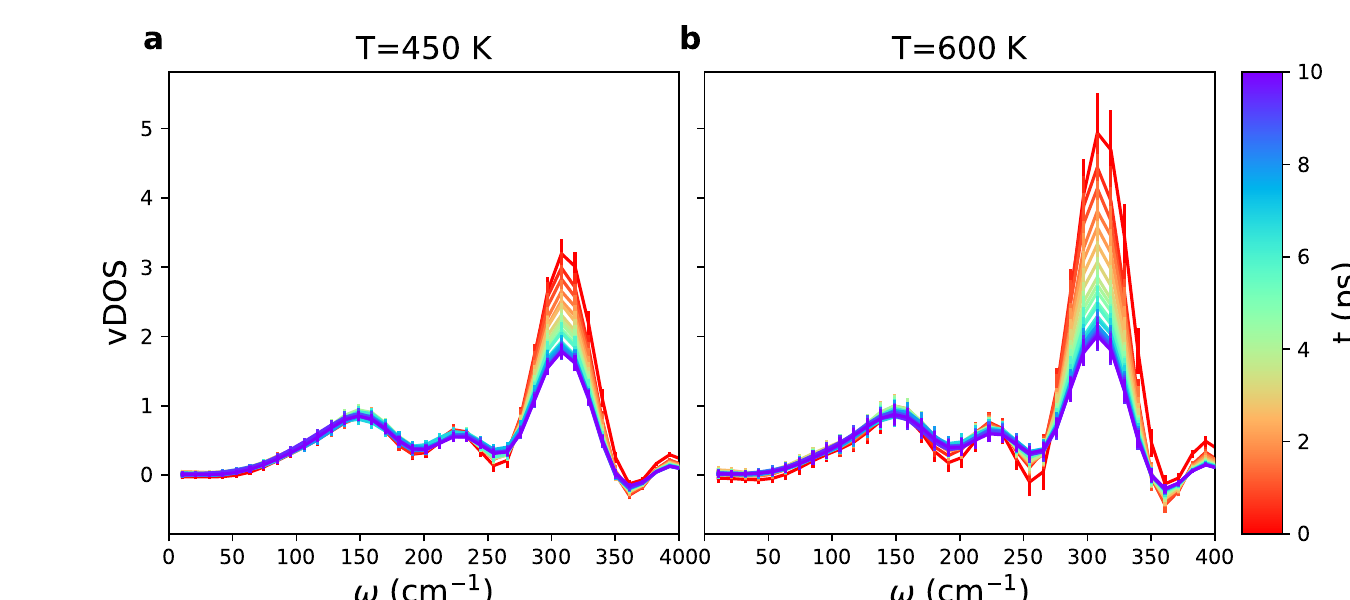}
\caption{Vibrational density of states during the relaxation as a function of time for (a) $T_{\rm exc}=450$~K, and\ (b) $600$~K}
\label{fig:vdos_time}
\end{center}
\end{figure}

Panel a shows the temperature rise during the 500~fs excitation for both cases, efficiently reaching $T_{\rm exc}=450$~K and $600$~K. In panel b we report the temperature difference during the relaxation phase: as anticipated in the main text, it follows a two-times exponential decay which we fit with the black-dashed curve. In particular, the fast decay occurs within $\sim 3$~ps reflecting the phonon thermalization dynamics. The second slower process suggest instead a spatial thermal diffusion from the locally heat area to the unperturbed regions.
\paragraph{Spectral features} 
As mentioned in the main text, the GLE inject energy in a specific frequency interval resulting in an enhanced intensity of the corresponding peak in the vDOS. During the NVE, this excess energy will be quickly redistributed within the vibrational modes of the excited region atoms, and this process will dominate the initial phase of the relaxation. A representation of this phenomenon is reported in \autoref{fig:vdos_time}

\begin{figure}[!h]
\begin{center}
\includegraphics[width=\textwidth]{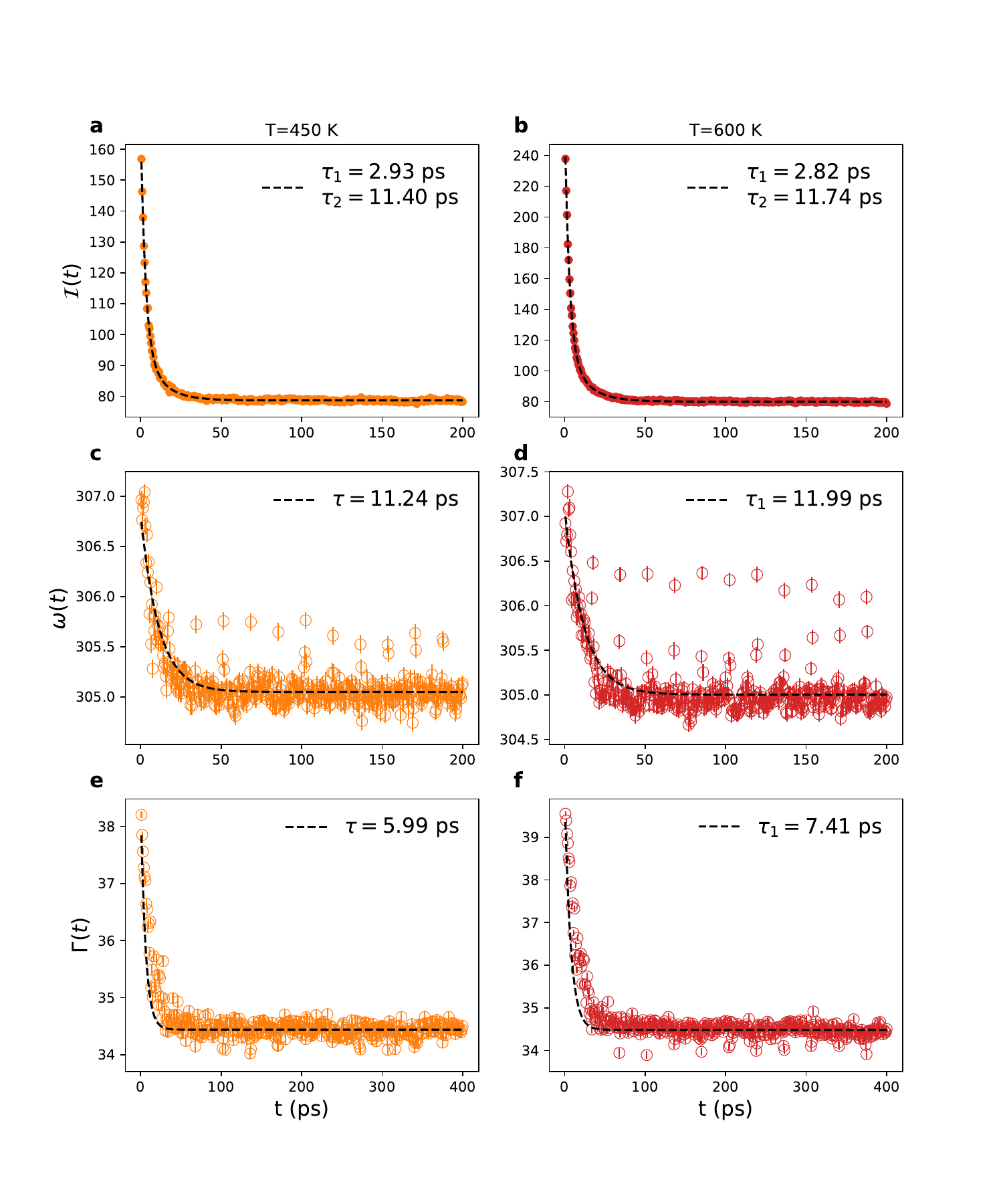}
\caption{Time dependence of the extracted spectral features for $T_{\rm exc}=450$~K (left column), and $600$~K (right column). a-b)~Excited peak intensity along with the two exponential model fit; c-d)~Excited peak frequency along with the single exponential model fit; e-f)~Excited peak FWHM along with the single exponential model fit.}
\label{fig:300_spectral}
\end{center}
\end{figure}

Consistent with the experimental analysis, we performed the extraction of the spectral features for the optical peak by considering the frequency interval between 260 and 340~cm$^{-1}$. In \autoref{fig:300_spectral}a and b we show the extracted intensity of the excited optical peak as a function of time, using the intensity equation in \autoref{eq:spectral_features}, and fitted using a two-times exponential decay

\begin{figure}[!h]
\begin{center}
\includegraphics[width=\textwidth]{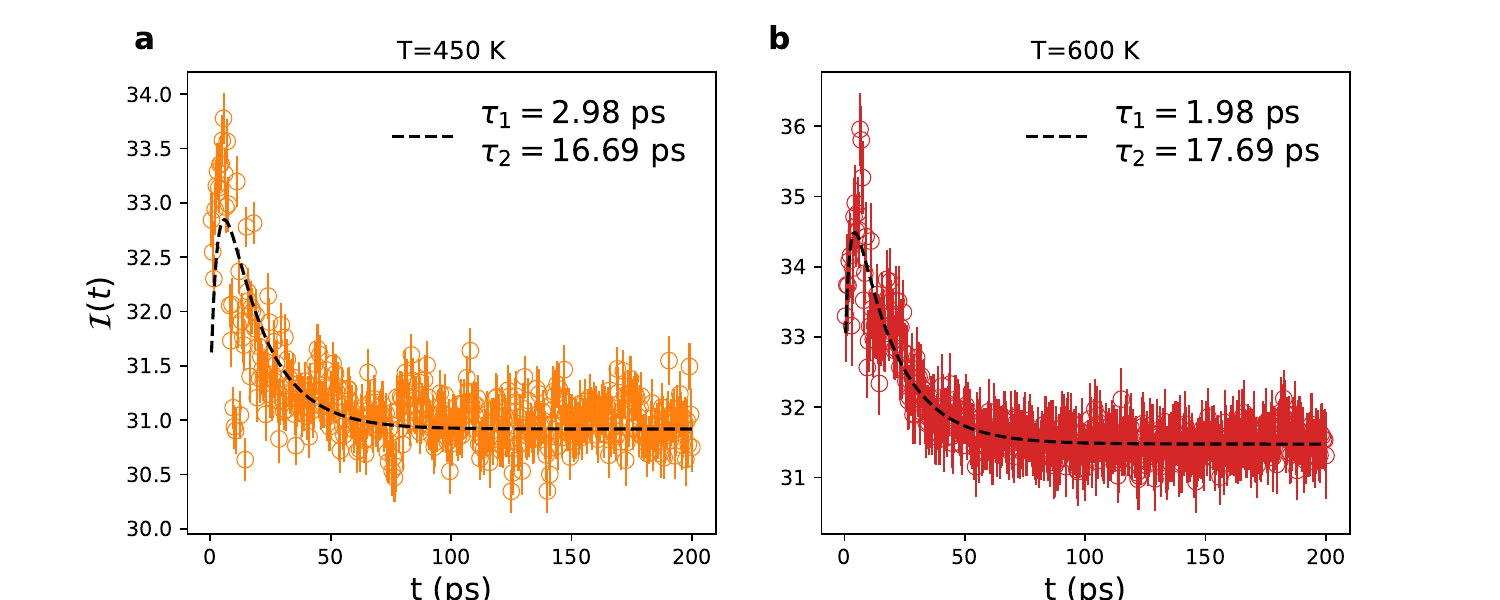}
\caption{Time dependence of the $220$~cm$^{-1}$ peak intensity along with the two exponential model fit for (a) $T_{\rm exc}=450$~K, and (b) $600$~K.}
\label{fig:220_area}
\end{center}
\end{figure}

\begin{figure}[!h]
\begin{center}
\includegraphics[width=\textwidth]{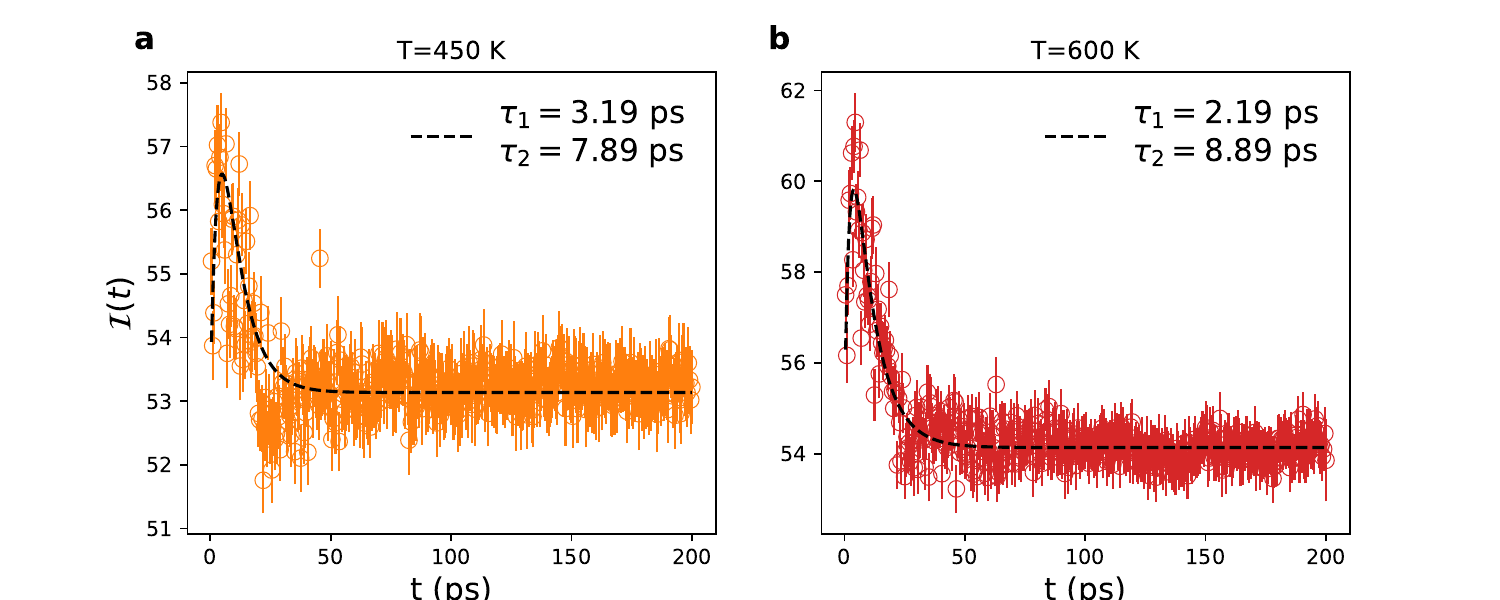}
\caption{Time dependence of the $150$~cm$^{-1}$ peak intensity along with the two exponential model fit for (a) $T_{\rm exc}=450$~K, and (b) $600$~K.}
\label{fig:150_area}
\end{center}
\end{figure}

Frequency shift was extracted by performing a weighted average as suggested in \autoref{eq:spectral_features}. As commented in the main text, the frequency decays following a single exponential and with a time scale that agrees with the spatial thermal diffusion identified in \autoref{fig:kinetic_temp}. Furthermore, we found that the frequency shift is convoluted with oscillations, as suggested by the presence of regularly spaced peaks in \autoref{fig:300_spectral}c and d. These are closely related to the thermal expansion caused by the local heat of the excitation, and will be commented more in detail when illustrating the atomic stress calculations. Finally, as mentioned in the main text, \autoref{fig:300_spectral}e and f show the FWHM of the excited peak as a function of time: the time evolution follows a single exponential decay, with a somehow intermediate timescale with respect to the peak intensity and the peak frequency: it results from an interplay between the strong anhamornicity induced by the excitation and the heating effect following the system relaxation.
To illustrate the phonon thermalization via mode coupling, we calculated the peak intensity for the mid-range frequency peak at $\sim220$~cm$^{-1}$ and for the acoustic region of the vibrational spectrum centered around $150$~cm$^{-1}$. As explained in the main text, these intensities decay with a two-times exponential decay but with a fast rise (within few ps) and a following slow decay (\autoref{fig:220_area} and \autoref{fig:150_area}). We infer that the fast rise is the fingerprint of phonon scattering originating from optical modes, while the slower component is again a reflection of the thermal diffusion following the initial relaxation.

These data are significantly noisier than the ones obtained from the optical peak: the peaks are broader and less intense, thus much more sensitive to the numerical uncertainty. In fact, the optical phonons will scatter with the lower frequency phonons of the atoms in the excited spot but also with the surrounding atoms from the unexcited regions. For this reason, the excess energy is transferred to a larger reservoir of atoms than the one we are using to extract the spectral features (that we remark being extracted by considering only the excited atoms). In other words, to obtain the peak intensity for the lower frequency modes there is less available vibrational energy per-atom.

\paragraph{Atomic stress}
As mentioned in the main text, an important role in the relaxation of the system is played by the thermal expansion caused by the local heat following the excitation. To quantify this aspect, we used the per-atom stress tensor as computed by LAMMPS. The tensor is computed using the virial theorem, thus accounting for both kinetic and potential contributions
\begin{equation}
    S_i^{\alpha\beta} = -\left( m_i v_i^\alpha v_i^\beta + \sum_{j \ne i} r_{ij}^\alpha F_{ij}^\beta \right)
\end{equation}
where $\alpha, \beta \in \{x,y,z\}$ are Cartesian components, $m_i$ is the mass of atom $i$, $v_i^\alpha$ is the $\alpha-$component of the velocity of atom $i$, $r_{ij}^\alpha$ is the $\alpha-$component of the displacement vector from atom $i$ to atom $j$, and $F_{ij}^\beta$ is $\beta-$component of the force exerted on atom $i$ by atom $j$. By convention, a positive stress means tension and a negative stress means compression. The atomic stress is averaged over all atoms in the two sub-regions to obtain an atomic stress difference as a function of time.

\begin{figure}[!h]
\begin{center}
\includegraphics[width=\textwidth]{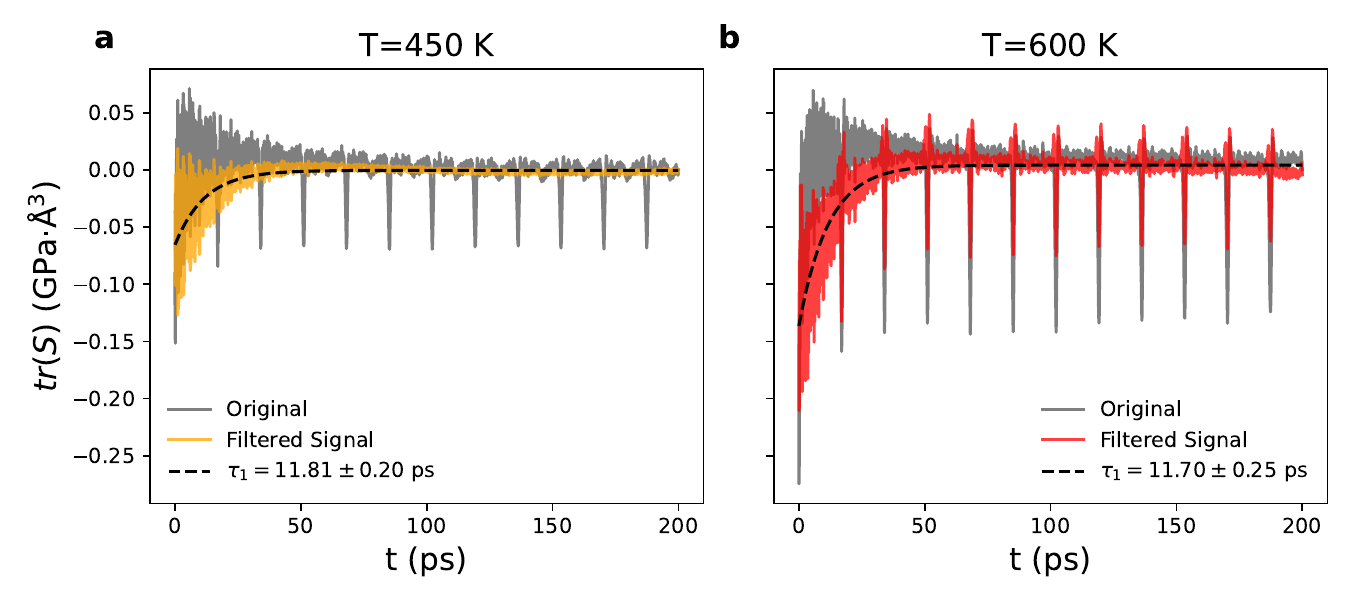}
\caption{Time dependence of the atomic stress difference between the excited and unexcited region (grey curve), the filtered data using the Savitzky-Golay filter (orange and red curves) along with the single exponential model fit for (a) $T_{\rm exc}=450$~K, and (b) $600$~K.}
\label{fig:stress}
\end{center}
\end{figure}

\autoref{fig:stress} shows the on-the-fly computed atomic stress during the NVE run (grey curves): it starts from a negative value (the excited region tends to expand due to the higher temperature) approaching zero as the equilibrium is reached. However, the raw data present some regular spikes which are the consequence of size effects: as a consequence of an initial thermal gradient and stress imbalance in the simulation cell, collective vibrational modes of atoms propagate mechanical stress through the lattice. Due to the finiteness of the simulation cell, they reflect off periodic images, and interfere constructively or destructively with ongoing stress oscillations. Since these spikes occurs approximately every 15 ps, using an average speed of sound for the Tersoff potential ($\sim 3800$~m/s) we get $v_s T/2\approx 2.8$~nm which is the transverse section of our sample, confirming the relation between these oscillations and the finiteness the simulation cell. These were filtered out using the Savitzky-Golay filter (although for $T=600$~K they could not be completely eliminated) and we fitted the resulting stress relaxation with a single exponential: the relaxation time of $11.8\pm0.2$ ps ($11.7\pm0.3$) for $T=450$~K ($T=600$~K), correspond to the time-scale observed in the frequency shift, hence confirming the role played by thermal expansion.

\subsubsection{Size effects}
\begin{figure}[!h]
\begin{center}
\includegraphics[width=\textwidth]{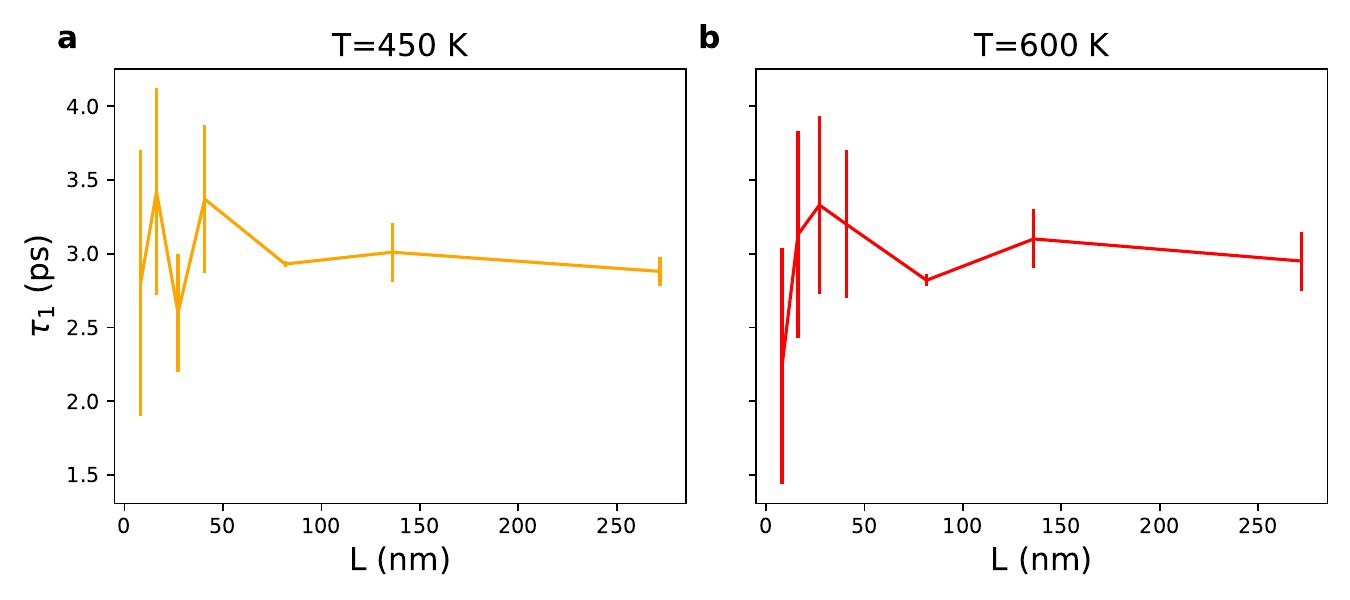}
\caption{$\tau_1$ calculated from the excited peak intensity decay as a function of system size for (a) $T_{\rm exc}=450$~K, and (b) $600$~K.}
\label{fig:size_effect}
\end{center}
\end{figure}

As mentioned in the previous section, non-equilibrium MD simulation are often affected by size-effect due to the unavoidable finiteness of simulation boxes. Depending on the observable different approaches can be adopted to address this issue, but it is fundamental to perform convergence tests. For this reason, we calculated the decay time of the excited optical peak as a function of the simulation cell length: we varied the the length keeping the width of the excited spot unchanged, basically varying the size ratio between the excited and unexcited region. In fact, in a laboratory setup the laser excites a region which is infinitely smaller than the sample size, which acts a infinite "reservoir". To understand what length scale allows a proper investigation of the relaxation dynamics, we report in \autoref{fig:size_effect} the fast decay time (identifying the phonon thermalization process) as a function of the cell length. In fact, as previously observed~\cite{DettoriJPCL19}, vibrational relaxation and thermal diffusion typically occur on different scales and the two process cannot be efficiently decoupled if not considering the proper distances.
 
From our tests we conclude that the $\tau_1$ decay time doesn't vary appreciably for $L>50$~nm, hence we considered simulation cells $82.1$~nm-long as a compromise between accuracy and computational workload. 
